\documentclass{aa}

\usepackage{amsmath}
\usepackage{amssymb}
\usepackage[varg]{txfonts}

\usepackage{graphicx}
\usepackage{natbib}
\usepackage{xcolor}
\usepackage[hidelinks]{hyperref}
\usepackage{placeins}
\graphicspath{{fig/}}
\usepackage{microtype}

\usepackage{orcidlink}

\bibpunct{(}{)}{;}{a}{}{,}

\let\tn\text
\let\vec\boldsymbol

\begin{document}

\title{Don't torque like that}
\subtitle{Measuring compact object magnetic fields with
analytic torque models}
\author{J.\,J.\,R.~ Stierhof\orcidlink{0000-0001-9537-7887}\inst{\ref{inst:remeis}}\thanks{\email{jakob.stierhof@fau.de}}
\and E.~Sokolova-Lapa\orcidlink{0000-0001-7948-0470}\inst{\ref{inst:remeis}}
\and K.~Berger\orcidlink{0009-0009-5344-9112}\inst{\ref{inst:remeis}}
\and G.~ Vasilopoulos\orcidlink{0000-0003-3902-3915}\inst{\ref{inst:uathens1},\ref{inst:uathens2}}
\and P.~Thalhammer\orcidlink{0000-0001-6269-2821}\inst{\ref{inst:remeis}}
\and N.~Zalot\orcidlink{0009-0000-3273-5659}\inst{\ref{inst:remeis}}
\and R.~Ballhausen\orcidlink{0000-0002-1118-8470}\inst{\ref{inst:umcp}, \ref{inst:gsfc}}
\and I.~El~Mellah\orcidlink{0000-0003-1075-0326}\inst{\ref{inst:usach}, \ref{inst:ciras}}
\and C.~Malacaria\orcidlink{0000-0002-0380-0041}\inst{\ref{inst:inaf}}
\and R.~E.~Rothschild\orcidlink{0000-0003-0396-0659}\inst{\ref{inst:ucalifornia}}
\and P.~Kretschmar\orcidlink{0000-0001-9840-2048}\inst{\ref{inst:esa}}
\and K.~Pottschmidt\orcidlink{0000-0002-4656-6881}\inst{\ref{inst:gsfc}, \ref{inst:umbc}}
\and J.~Wilms\orcidlink{0000-0003-2065-5410}\inst{\ref{inst:remeis}}
}
\institute{Dr.~Karl~Remeis-Sternwarte and Erlangen Centre for
  Astroparticle Physics, Friedrich-Alexander-Universit\"at
  Erlangen-N\"urnberg, Sternwartstr.~7, 96049 Bamberg, Germany\label{inst:remeis}
\and
Department of Physics, National and Kapodistrian University of Athens, University Campus Zografos, GR 15784, Athens, Greece\label{inst:uathens1}
\and
Institute of Accelerating Systems \& Applications, University Campus Zografos, Athens, Greece\label{inst:uathens2}
\and
University of Maryland College Park, Department of Astronomy, College Park, MD 20742, USA\label{inst:umcp}
\and
NASA Goddard Space Flight Center, Astrophysics Science Division, 8800 Greenbelt Road, Greenbelt, MD 20771, USA\label{inst:gsfc}
\and
Departamento de Física, Universidad de Santiago de Chile, Av. Victor Jara 3659, Santiago, Chile\label{inst:usach}
\and
Center for Interdisciplinary Research in Astrophysics and Space Exploration (CIRAS), USACH, Santiago, Chile\label{inst:ciras}
\and
INAF-Osservatorio Astronomico di Roma, Via Frascati 33, I-00076, Monte Porzio Catone (RM), Italy \label{inst:inaf}
\and
Department of Astronomy and Astrophysics, University of California San Diego, 9500 Gilman Dr., La Jolla, CA 92093-0424,
USA\label{inst:ucalifornia}
\and
European Space Agency (ESA), European Space Astronomy Centre (ESAC),
Camino Bajo del Castillo s/n, 28692 Villanueva de la Cañada, Madrid,
Spain \label{inst:esa}
\and
CRESST and Center for Space Sciences and Technology, University of
Maryland Baltimore County, 1000 Hilltop Circle, Baltimore, MD 21250,
USA\label{inst:umbc}
}
\authorrunning{J.~J.~R.~Stierhof et al.}

\abstract{Changes of the rotational period observed in various magnetized accreting sources are generally attributed to the interaction between the in-falling plasma and the large-scale magnetic field of the accretor. A number of models have been proposed to link these changes to the mass accretion rate, based on different assumptions on the relevant physical processes and system parameters. For X-ray binaries with neutron stars, with the help of precise measurements of the spin periods provided by current instrumentation, these models render a way to infer such parameters as the strength of the dipolar field and a distance to the system. Often, the obtained magnetic field strength values contradict those from other methods used to obtain magnetic field estimates.}
{We want to compare the results of several of the proposed accretion models. To this end an example application of these models to data is performed.}
{We reformulate the set of disk accretion torque models in a way that their parametrization are directly comparable. The application of the reformulated models is discussed and demonstrated using Fermi/GBM and Swift/BAT monitoring data covering several X-ray outbursts of the accreting pulsar 4U\,0115$+$63.}
{We find that most of the models under consideration are able to describe the observations to a high degree of accuracy and with little indication for one model being preferred over the others. Yet, derived parameters from those models show a large spread. Specifically the magnetic field strength ranges over one order
of magnitude for the different models. This indicates that the results are heavily influenced by systematic uncertainties.}
{The application of torque models provides a generic way to access system parameters of the accreting object. Values obtained via those models must be treated with caution, since the systematics of the models must be taken into account. Our example suggests that the current state of analytic torque models does not allow for quantitative measurements of the magnetic field of an accreting object. Systematic application to a sample of sources with known magnetic fields and distances will provide a selection criterion between models in the future.}


\date{Received/Accepted}

\maketitle

\section{Introduction}
\label{sec:intro}

The accretion of matter onto a compact object is one of the most
effective processes to convert gravitational potential energy into
radiation \citep[e.g.,][]{book:frank2002a}. There are a number of
systems for which accretion shapes the observed behavior, ranging from
accretion flows onto compact objects such as black holes (BHs),
neutron stars (NSs) and white dwarfs (WDs), to the evolution of T\,Tauri stars
\citep{hartmann1999a,mckee2007a}. In all of these
objects, the conditions are such that large amounts of matter are
captured by the gravity of the central object (``accretor'') at a high rate. For
stellar mass BHs, and for NSs and WDs, accretion primarily occurs in
binary systems where the source of the accreted matter is a companion
star (``donor''). In the case of T\,Tauri stars and for supermassive
BHs, the accreted matter is provided by a surrounding gas cloud
\citep[and references therein]{larson2003a}.

The accretion process and subsequent change of angular momentum of the
accretor was extensively discussed in the context of NS binary systems
in the last 50 years. Recently also other systems were discussed in a
very similar context. Analytical models that were derived using simplifying
assumptions for an accreting NS are, in fact, so minimal that many accreting
systems can be described within this picture. Due to the historical development
a certain focus on NSs cannot be avoided. It should be clear from the
descriptions of the models, however, that as long as the accreting object
can be regarded as a rotating magnetic dipole they should fall within the
scope of this work. The accretion process and the dynamics in the outer region of the
accretion stream are mainly shaped by gravity. Provided the magnetic field of
the accretor is strong enough the interaction between the rotating magnetic
field and the accreted plasma in the inner flow can significantly
affect the dynamics and channel the flow onto the accretor's magnetic
poles
\citep[e.g.,][]{elsner1977a,illarionov1975a,davidson1973a}\footnote{For
accretion flows around black holes, the magnetic field advected
inwards by the accreted material can also reach values that are high
enough that it plays an important role in the flow dynamics
\citep{uzdensky2004a, uzdensky2005a,nathanail2014a}, but we will
exclude these effects in this paper.}. Close to the poles, the
in-falling matter is decelerated, converting the high kinetic energy
to radiation \citep[][and references therein]{book:lipunov1992a}. The
details of the deceleration and emission process vary for different
mass accretion rates and are still a topic of active research
\citep[e.g.,][]{becker2007a,becker2012a,postnov2015a,
  farinelli2016a,west2017a,sokolovalapa2021a}. There is consensus in
the literature that the characteristic X-ray spectrum observed during
episodes of accretion encodes information about the plasma conditions
in the region where the radiation is produced. One of the crucial
parameters is the strength of the magnetic field which sometimes
imprints one or several localized cyclotron resonance scattering
feature (CRSFs, also known as ``cyclotron line'') onto the broad band
spectral continuum. These absorption features allow for a direct
measurement of the magnetic field strength at the scattering site. To
date, the magnetic fields of about 30 accreting NSs have been measured
with this method \citep[see][for a recent review] {staubert2019a}.

Not all observed spectra from accreting and strongly magnetized
neutron stars show a clear CRSF signal, however. For some sources this
may be due to the line being located outside of the observable energy
range, while for other sources it is possible that the line is weak or
suppressed. At the same time measurements of the magnetic field
strength are crucial for our understanding of the evolution and
development of such binary systems. Several indirect methods to assess
the magnetic field strength of NS accretors have been developed. For
example, the transition to radiation-dominated deceleration of the
falling plasma in the accretion column is expected to occur at a
luminosity level that depends on the magnetic field. It is further
generally assumed that the emission characteristic is different in the
case where a radiation shock has formed, compared to when the plasma
directly impacts the NS surface. Since this transition should be accompanied
by observable changes in spectral behavior, the magnetic field then
can be estimated based on measurements of this critical luminosity
\citep{becker2012a,mushtukov2015a}.

Other methods to determine the central magnetic field are based on
signatures of the interaction of the magnetic field with the accretion
stream where the flow is significantly influenced by the rotating
field \citep{davidson1973a}. Above a critical strength of the magnetic
field, plasma can no longer move freely but couples to the magnetic
field lines \citep{elsner1977a}. In this region, the specific
angular momentum of the plasma is fully determined by the rotation
rate and the distance to the rotation center, and therefore not
Keplerian, while far away from the accreting object the angular
momentum of the plasma is determined by the accretion mechanism. As a
consequence, a transition region forms, where the plasma can lose or gain
angular momentum through the interactions of the
accreting plasma with the magnetic field. This interaction causes a change
in the rotation of the accretor. Since this change of the spin
period can be observed in accreting neutron stars \citep[for a recent
  review]{malacaria2020a}, it is in principle possible to deduce the
accretor's magnetic field from such measurements.

Over the past 30\,years, a number of different torque models linking
the mass-accretion rate with the spin period have been proposed and
used to estimate the central magnetic field. Direct comparison of the
models is difficult, however, and we will show how derived $B$-field strength values
differ depending on the used model. Furthermore, inconsistent
terminology and model descriptions make it difficult to judge the
differences between them. In this paper, we reformulate the proposed
analytical models into a comparable form, in order to investigate their
similarities and differences. In Sect.~\ref{sec:accretion-physics} the
physics of the general accretion process is depicted and distinct
possible scenarios for torquing are discussed. In
Sect.~\ref{sec:torque-models} we focus on the disk accretion scenario
and review a number of proposed disk accretion torque models. We then
propose a convenient generalized parametrization to compare the models
with data in Sect.~\ref{sec:torque-application}. In
Sect.~\ref{sec:example} we describe the steps to apply those models to
data, providing an example for the Be X-ray binary 4U\,0115$+$63.
Section~\ref{sec:other-aspects} provides a discussion of additional
physical aspects, which are not included in the standard disk
accretion models, and other accretion scenarios. Conclusions are given
in Sect.~\ref{sec:conclusion}.

\section{From donor to accretor}
\label{sec:accretion-physics}

We start setting the scene by a discussion of the general properties
of the accretion flow. There are four scenarios envisioned how
matter is transported towards the accretor and captured in its gravitational field:
accretion from a wind, Roche lobe overflow, accretion from a
decretion disk, and accretion by magnetic coupling. The latter
scenario mainly applies to strongly magnetized WDs (polars) with the
largest magnetic dipole moments \citep[on the order of
  $10^{35}\,\mathrm{G}\,\mathrm{cm}^3$,][]{cropper1990a}, where the
strong magnetic field captures matter close to the inner Lagrange
point. This capture links the donor and accretor via the magnetic field and
forces the binary system into synchronous rotation, where the accretor
is magnetically locked to the donor up to the point where the strong
magnetic field extracts material from the surface of the companion
star \citep[][and references therein]{mukai2017a}. In this case the
spin of the accretor is directly linked to the orbital period and not
influenced by the accretion flow, such that this scenario is not of
interest here.

In the other three scenarios the matter is gravitationally captured further away from the accretor where the magnetic field is not yet dominating the plasma flow.
In compact systems where accretion happens via Roche lobe overflow, the donor extends
close to its Roche lobe such that material can fall into the
gravitational well of the accretor. Depending on the eccentricity of
the orbit, either steady accretion or periodic episodes of accretion
can occur. Since the matter transfer predominantly happens through the
inner Lagrange point, the flow will have significant angular momentum
relative to the accretor, leading to the formation of an accretion
disk \citep{pringle1972a, book:wheeler1993a}.

In wind-fed systems the companion star has a strong stellar wind that
can transport significant amounts of matter to the vicinity of the
accretor, where it is partly accreted. Whether an accretion disk forms
is subject to the residual angular momentum of the accreted matter.
The first description of accretion from a wind in high-mass X-ray
binaries was given by \citet{davidson1973a}. This description was
based on the model of \citet{bondi1944a} where exact cancellation of
angular momentum is assumed and therefore no disk formation is
possible. In a realistic system, however, even though a disk might not
form, it is still possible that the accreted matter carries
significant angular momentum inward \citep{hayasaki2004a,blondin2013a,
  elmellah2019a,elmellah2019b}.

We consider the accretion flow in systems where the matter is supplied from
a decretion disk as a separate case, although it exhibits to some
degree a mixture of the phenomena seen in wind-fed and Roche lobe overflow systems.
Such decretion disks occur in Be star systems, which are fast rotating stars
with outflows in their equatorial regions which can be detected by the
strong emission lines emitted from those disks. The mechanism that causes
the fast rotation is still under debate and possibly involves several
mechanism, such as binary interactions or pulsations of the star
\citep[see, e.g.,][] {rivinius2013a}. Isolated Be stars tend to have
very large disks, up to several hundred star radii \citep[see][for
  recent measurements]{klement2017a}. In a binary system, this disk is
truncated due to the tidal interaction slightly inside the Roche-lobe
of the Be star \citep{okazaki2002a,panoglou2016a}. The excess material of the
decretion disk of the companion Be star is then either captured directly by the
accretor, usually with large angular momentum causing the formation of an accretion disk, or
flows into the environment of the binary system. The compact objects in these
systems therefore experience (often periodic) episodes of strong accretion
which are observed as outbursts and are accompanied by very clear spin up
signals.

\subsection{Feeding the spinning top}
\label{sec:feeding}

From here on we assume that material is captured inside the
gravitational well of the accretor with sufficient angular momentum
such that accretion from a disk is the appropriate picture for
describing the accretion flow. In this simplified picture, the further
behavior of the accretion flow deeper in the potential well mainly
depends on the hierarchical order of some characteristic radii. We
assume that the material in the disk follows Keplerian orbits. Without
interactions with the central magnetic field the plasma can only transport
angular momentum outwards through diffusive processes or lose it via
disk winds \citep{book:frank2002a}.

\begin{figure}
\resizebox{\hsize}{!}{\includegraphics{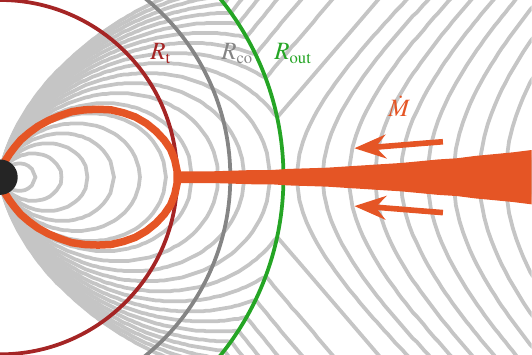}}
\caption{Sketch of the disk and magnetic field configuration for an
  accreting system (not to scale). The truncation radius $R_\tn{t}$
  indicates where the matter, indicated by the mass accretion rate,
  $\dot{M}$, couples to the magnetic field. The co-rotation radius
  where the magnetic field lines rotate with the same speed as the
  local Keplerian speed is at $R_\tn{co}$. The outer radius,
  $R_\tn{out}$, indicates the outermost radius where the central
  magnetic field is connected to the disk. Beyond this radius the
  stress becomes too large, causing the field to disconnect from the
  central source.}
\label{fig:sketch}
\end{figure}

The orbits of the matter in the disk can be divided into two regions,
depending on whether their orbital (Keplerian) frequency is smaller or
larger than the spin frequency of the accretor (see
Fig.~\ref{fig:sketch}). The transition between the regions occurs at
the \emph{co-rotation radius},
\begin{equation}\label{eq:rco}
R_\tn{co} = \left(\frac{G M}{\Omega^2}\right)^{1/3}\,,
\end{equation}
where $G$ is the gravitational constant, $M$ is the mass of the
accretor, and $\Omega$ is the accretor's spin frequency.

A second important radius defines the region in the accretion flow
where the plasma couples to the magnetic field. This region is usually
defined as the volume where the energy density of the in-falling
plasma is smaller than the energy density of the magnetic field. The
boundary where the energy densities are equal is called \emph{Alfv\'en
surface} \citep{alfven1947a,lamb1973a}. This surface has generally a
complicated shape, depending on the structure of the surrounding
plasma and the magnetic field. Its approximate shape is often derived
by assuming a spherically symmetric plasma distribution that is only
accelerated towards the gravitational center and by describing the
central magnetic field with a dipole configuration with dipole moment
$\mu$. The field strength at point $\vec{r}$ is then given by
\begin{equation}
B^2(\vec{r}) = \frac{\mu^2}{r^6}(3\cos^2\theta + 1)\,,
\end{equation}
where $\theta$ is the angle between $\vec{r}$ and the direction of
the dipole moment (specifically, $B = 2\mu/r^3$ in the direction of
the dipole moment). In this case the equilibrium condition gives
the \emph{Alfv\'en radius}
\begin{equation}
\label{eq:ralfven}
R_\tn{A} = \left(\frac{\mu^4}{2GM\dot{M}^2}\right)^{1/7}\,,
\end{equation}
where $\dot{M}$ is the mass accretion rate and where a term that
depends on the angle between the dipole moment and the radius vector
is ignored. The difference between $R_\tn{A}$ and the true border of
the Alfv\'en surface is at most a factor of 2
\citep[e.g.,][]{book:campbell2018a}.

Note that the term \emph{Alfv\'en radius} and its corresponding symbol
are often used as a characteristic radius defining regions in the
accretion flow as well as for the Alfv\'en surface (that is, the
surface where the equilibrium condition is fulfilled), without a clear
specification of any assumptions regarding the plasma and magnetic
field. In this work we exclusively use it for the expression given in
Eq.~(\ref{eq:ralfven}).

In all torque models discussed in the following, it is furthermore
assumed that a radius $R_\tn{t}$ exists at which the accretion disk is
truncated because the matter is forced to follow the magnetic field
lines. At this point the accretion flow transits from Kepler orbits to
co-rotation and then falls along the magnetic field lines onto the
surface of the accretor \citep[see,][]{pringle1972a,
  lamb1973a,basko1976b,ghosh1977a,popham1991a}. In the limit of a low
magnetic field, where $R_\tn{t}$ is located at the surface or inside
the accretor, angular momentum can be exchanged through friction,
providing a limit to the angular momentum exchange \citep[see,
  e.g.,][p.~153ff.]{book:frank2002a,syunyaev1986a}.

In the most simplified model for torqued compact objects it is assumed
that the difference is fully transferred to the accretor
\citep{pringle1972a,ghosh1979b}. Describing the accretion disk with an
$\alpha$-disk model \citep{shakura1973a}, \citet{ghosh1979b} found
that for $R_\tn{t} \ll R_\tn{co}$ the transition radius scales with
the Alfv\'en radius, and that $R_\tn{t}/R_\tn{A} \sim 0.52$ for
typical NS accretors. Any difference of the angular momentum between
the matter at $R_\tn{t}$ and at the surface of the compact object must
be dissipated to enable accretion. Whether the total torque onto the
accretor is positive or negative then depends on the interaction
between the magnetic field and the accretion disk. \citet{ghosh1979b}
assumed that a large part of the disk is linked to the accretor's
magnetic field such that the part of the disk inside $R_\tn{co}$, and
therefore rotating faster than the field, exerts a positive torque on
the central object. The part outside $R_\tn{co}$, on the other hand,
extracts torque from the accretor.

More recent models for torquing link the radius up to which the disk
is connected, $R_\tn{out}$, to the shear of the field inside the disk
instead, no matter the exact location of $R_\tn{t}$. This shear is
commonly expressed as a function of the pitch angle of the field, up
to a maximum allowed pitch which is related to the parameters of the
disk's plasma (viscosity, diffusivity) and is generally assumed to be
of order unity \citep[see, e.g.,][]{wang1995a,matt2005a}.

Numerical simulations have shown that the physical process of the
coupling of the accretion flow to the magnetic field is more complex
than the simple picture drawn so far \citep[e.g.,][]{romanova2008a}.
Due to the coupling of the plasma at $R_\tn{t}$ a current is induced
in the disk that causes a counter field opening up the magnetic field
lines at $R_\tn{t}$ (i.e., $R_\tn{out} \approx R_\tn{t}$). The central
field and disk are therefore only barely directly connected. However,
the opening of the field at $R_\tn{t}$ causes re-connections of the
field, driving out plasma as magnetospheric ejections
\citep[e.g.,][]{zanni2013a}. These ejections are trapped in the sheath
layer between the central field and the counter field of the disk.
The field of the disk and the star are connected to the magnetospheric ejections, such that they
provide an indirect link between the central field and the disk,
resulting in an effective outer radius, $R_\tn{out}$
\citep{ireland2020a,ireland2022a}\footnote{The MHD simulations cited
here were conducted in the context of T\,Tauri stars. These have
central magnetic fields several orders of magnitude lower than
accreting neutron stars. Nevertheless, it is expected that the
magnetic field at $R_\tn{t}$ acts mostly as a scaling factor and so
overall results should still hold.}.

In Sect.~\ref{sec:torque-models} we describe several analytical
models and explicitly give the predicted torques. As discussed above,
the spin up due to accretion depends on the specific angular momentum
at the inner disk edge of the disk, and therefore on the location of
$R_\tn{t}$, while the differential torque between disk and central
field depends on the physics of the connected region. In all models we
reviewed the torque is derived from specific assumptions on $R_\tn{t}$
and $R_\tn{co}$. Independent of the details of the specific model, the
total torque exerted by the accreting matter on the central object can
then be written as
\begin{equation}
\label{eq:torquefun}
\tau = \tau_\tn{t} + \tau_\tn{m} = \tau_\tn{t} n(\omega)\,.
\end{equation}
where $\tau_\tn{m}$ is the contribution due to the magnetic
interaction (differential rotation between disk and magnetic field)
and where
\begin{equation}
\label{eq:keplertorque}
\tau_\tn{t} = \dot{M}\sqrt{G M R_\tn{t}}\,.
\end{equation}
is the torque that corresponds to the angular momentum at $R_\tn{t}$
in the accretion disk plane \citep{ghosh1979b}. \citet{elsner1977a}
show that by factoring out $\tau_\tn{t}$ the dimensionless function
$n$ defined in Eq.~(\ref{eq:torquefun}) only depends on one parameter,
which is called the \emph{fastness}
\begin{equation}
\label{eq:fastness}
\omega = \frac{\Omega}{\Omega_\tn{K}(R_\tn{t})} =
\left(\frac{R_\tn{t}}{R_\tn{co}}\right)^{3/2}\,,
\end{equation}
where $\Omega_\tn{K}(r)$ is the Kepler orbit frequency at
radius $r$,
\begin{equation}
\label{eq:omega_k}
\Omega_\tn{K}(r) = \sqrt{\frac{GM}{r^3}}\,.
\end{equation}

\subsection{Dietary restrictions}
\label{sec:hunger}

Since the accretion behavior of NSs was discovered in the early 1970s,
most observations of these sources were done during bright states,
that is, at times of large mass accretion rates. It is important,
however, to also consider accretion flow at lower mass accretion.
Originally, it was believed that when not actively accreting, the fast
rotating magnetic field of NS prevents matter from getting closer than
a certain radius \citep{illarionov1975a,campana2001a}. This \emph{propeller state} is
believed to be entered when $\omega > 1$. \citet{perna2006a} showed
with a simple ballistic calculation, not taking dissipative processes
of plasma or gas dynamics into account, that one can obtain a critical
radius, $R_\tn{prop}$, and a corresponding fastness,
$\omega_\tn{prop}$, such that if $R_\tn{t} \gtrsim R_\tn{prop}$ matter
is ejected from the system. They also demonstrated that there is a
truncation radius less than $R_\tn{prop}$, corresponding, however, to
$\omega > 1$, such that the matter is trapped between the
gravitational potential and the rotating magnetic barrier. It is not
clear if in this case accretion proceeds normally. There are
arguments, however, that in this case the disk is truncated at
$R_\tn{co}$, accretion is greatly reduced and unstable
\citep{rappaport2004a,romanova2008a,romanova2018a}.

Three other aspects can also prevent matter from getting closer to the
central object and under the influence of the magnetic field. Those are
related each to a specific characteristic radius that gives the size of the
accreting system. First, in wind-fed systems matter is efficiently accreted only if it is
inside the \emph{accretion radius} (often also called the \emph{gravitational capture radius})
\begin{equation}
R_\tn{ac} = \frac{2GM}{v^2}
\end{equation}
obtained from simple ballistic calculation of a particle with velocity
$v$ \citep[e.g.,][]{edgar2004a}. The extent of the capture radius
(subject to the wind properties) determines how much material is
available for accretion.

Secondly, in fast-rotating pulsars with periods of milliseconds,
typical for accreting millisecond pulsars (MSPs), a strong pulsar wind
is produced by the rotating magnetic field. The characteristic radius
determining the pulsar wind is the \emph{light cylinder}
\begin{equation}
R_\tn{lc} = \left(\frac{2R_\tn{co}^3}{R_\tn{S}}\right)^{1/2}
\end{equation}
where $R_\tn{S} = 2GM/c^2$ is the \emph{Schwarzschild radius} and $c$
is the speed of light. During the accretion process the disk models
are expected to hold with a reduced efficiency due to the pressure of
the pulsar wind, up to a point where the pulsar wind can prevent
accretion altogether \citep{shvartsman1970a,parfrey2016a}.

Finally, we note that besides pressure due to the rotating
magnetosphere, the static magnetic field pressure alone can (depending
on the conditions of the plasma) also prevent or significantly reduce
accretion \citep{toropina2003a}. This is particularly relevant for old neutron
stars accreting from the interstellar medium (as opposed to a companion) but might
also have an effect for wind accreting systems with conditions such that the ram
pressure is relatively low. A review of the different stages for
magneto-rotating accretors can be found in
\citet{lipunov1987a} and in \citet[however, see also
  \citealp{phd:sokolovalapa2023a}]{bozzo2008a}.

\section{Different ways to connect plasma}
\label{sec:torque-models}

From the previous description and assuming that the magnetic field structure is a dipole, \citet{elsner1977a} and \citet{ghosh1979b} established that episodes
of stable accretion are only possible when $\omega < 1$, that is,
$R_\tn{t} < R_\tn{co}$. The change of the angular frequency of the
central object, $\tn{d}\Omega$ is related to the total torque via the
change in angular momentum, $\tn{d}\mathcal{L}$, and moment of inertia, $I$
\begin{equation}
\tau = \frac{\tn{d}\mathcal{L}}{\tn{d}t} = I
\frac{\tn{d}\Omega}{\tn{d}t}\,.
\end{equation}
Here it is assumed that the moment of inertia is constant. Substituting
the angular frequency with the spin period, $P$, 
\begin{equation}
\label{eq:spin-up}
-2\pi I \dot{P} = \tau_\tn{t} n(\omega) P^2\,,
\end{equation}
where $\tau$ is expressed using Eq.~(\ref{eq:torquefun}) and $\dot{P}=dP/dt$.

Equation~(\ref{eq:spin-up}) can then be used to estimate the total torque exerted on the
central object provided measurements of $P$, $\dot{P}$, and reasonable
estimates of $I$. It is necessary to determine $\omega$ to calculate the change in the
period. The fastness and $n(\omega)$ are tightly linked and therefore have to be determined
together. In Sect.~\ref{sec:rundown}, we will summarize and compare some of the
common descriptions for this quantities.

\subsection{One description to torque them all}
\label{sec:torque-application}

\begin{figure}
\resizebox{\hsize}{!}{\includegraphics{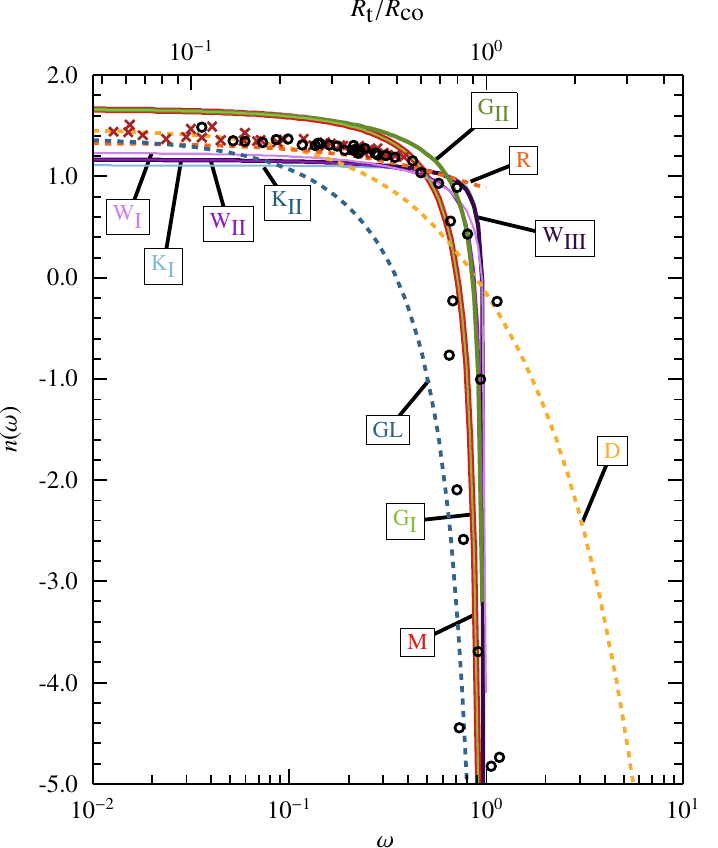}}
\caption{Torque function, $n(\omega)$, as a function of the model-dependent fastness, $\omega$, for the torque
  models discussed in Sect.~\ref{sec:rundown}. Dashed lines correspond to
  models where $R_\tn{t} \propto R_\tn{A}$ is assumed. Red crosses
  show the values obtained by \citet{ireland2020a}, black circles are
  from \citet{ireland2022a}, see Table~\ref{tab:symbols} for an explanation
  of the labels. As discussed in the text, no conclusions concerning the
  relative behavior of $n(\omega)$ between different models should be
  drawn from this figure.}
\label{fig:torque-default}
\end{figure}

Before we discuss specific model descriptions in Sect.~\ref{sec:rundown}, we have to address a subtle property of torquing models that complicates their interpretation. In general, torquing models 
can be roughly separated into two families: one where the approximation of
\citet{ghosh1979b} is assumed to hold and a matching function $n$ is
constructed \citep[e.g.,][]{dai2006a}, and one where the plasma and magnetic
field interaction and geometry is assumed from which $n(\omega)$,
is derived \citep[e.g.,][]{wang1995a,matt2005a}. In the
latter case, it turns out, that $\omega \propto R_\tn{A}^{3/2}$ (from Eq.~\ref{eq:fastness}
and $R_\tn{t} \propto R_\tn{A}$) does not
hold in general as the approximation of \citet{ghosh1979b} is only valid
in a small range and for $\omega \ll 1$. 
Since the fastness is defined via the truncation and the co-rotation radii, it is an implicit function of the assumed magnetic field geometry and the coupling physics. This means that the fastness parameter depends on the specific assumptions of a given physical model. In particular, a value of $\omega$ is realized for different values of $\mu$ and $\dot{M}$ for different models. This degenerate situation makes it difficult to compare models when expressed as a function of $\omega$ only. For example, from the comparison of different $n(\omega)$ shown in Fig.~\ref{fig:torque-default} one could get the impression that the model of \citet{ghosh1979b} deviates significantly from the majority of other torquing models. As we shall show below, however, most of the deviation is due to the fact that $\omega$ has a specific meaning only within a given model. From here on $\omega$ will be replaced by $\omega_i$ to reflect this in notation, where $i$ is a label referring to a specific model

In order to enable a direct comparison between different models, we therefore introduce a
unifying description to treat all considered models in the same way, using well-defined physical quantities that are independent of the particular assumptions of the models. To this effect we introduce the \emph{canonical fastness}
\begin{equation}
\hat\omega = \left(\frac{R_\tn{A}}{R_\tn{co}}\right)^{3/2}\,
\end{equation}
and express all considered torque functions with this variable. 

\subsection{Model rundown}\label{sec:rundown}

\begin{table*}
\renewcommand{\arraystretch}{1.05}
\caption{List of torque models and their symbols}
\label{tab:symbols}
\begin{tabular*}{\hsize}{@{\extracolsep{\fill}}lll}
\hline
Reference & Labels & Description\\
\hline
\citet{ghosh1979b} & GL & ``standard'' model derived for a number of simplifying approximations \\
\citet{wang1995a} & W$_\tn{I,II,III}$ & three models derived based on different transport mechanisms \\
\citet{rappaport2004a} & R & estimation of disk truncation for fast systems \\
\citet{matt2005a} & M & fully parametrized model coupling strength \\
\citet{dai2006a} & D & ad-hoc model to extend truncation beyond co-rotation radius \\
\citet{kluzniak2007a} & K$_\tn{I,II}$ & similar to \citeauthor{wang1995a}, solves disk structure \\
\citet{gao2021a} & G$_\tn{I,II}$ & extension of \citeauthor{wang1995a} for stronger coupling \\
\hline
\end{tabular*}
\tablefoot{The numerals in the subscript denote the different
  models discussed in the respective publications.
}
\end{table*}

We will now  give an overview of several analytic torque models that
have been proposed in the literature. The selection of the models does not aim for
completion nor are any criteria applied, but 
it is based on an initial search for different approaches that all could be
rewritten in a form using the unification as discussed above. In general, we will give $n(\omega_i)$ using a consistent notation and then show how the model-specific $\omega_i$ can be derived for a given (model-independent) $\hat\omega$. Table~\ref{tab:symbols} gives an overview of the models addressed in the following.
We note that 
several other
models have been proposed that cannot be described easily in this way
and therefore had to be excluded (see Sect.~\ref{sec:quasisphere}
for an example). 

Probably the most influential model for torquing was described
by \citet{ghosh1979b}. As mentioned earlier, in the analytic form
of the model $R_\tn{t} \propto R_\tn{A}$ is adopted.
\citeauthor{ghosh1979b} assumed that the accretor's magnetic field
causes a response field in part of the $\alpha$-disk. Further
assumption of the structure of this field led to an analytic
approximation of the dimensionless torque function
\begin{equation}
n_\tn{GL}(\omega_\tn{GL}) = 1 +
\frac{2.22\omega_\tn{GL}-5.60\omega_\tn{GL}(1-\omega_\tn{GL})^{0.173}+0.39}{1-\omega_\tn{GL}}\,,
\end{equation}
which is claimed to be valid for $\omega < 0.9$
\citep[Eq.~10]{ghosh1979b}. The restriction in $\omega$ here is only due
to the approximation of the derived integral and not related to the
estimate in the magnetospheric radius.
The linear relation derived in the model is generally expressed as
\begin{equation}
\label{eq:rt-linear-fix}
\omega_\tn{GL} = \Gamma^{3/2}\hat\omega\,.
\end{equation}
with $\Gamma = 0.52$ for \citet{ghosh1979b}. While this directly allows
one to express $R_\tn{t} = \Gamma R_\tn{A}$, also called the 
\emph{magnetospheric radius}, it will become clear that it is only valid
in a narrow range.

\citet{wang1987a} showed that the field configuration assumed by
\citet{ghosh1979b} overestimated the toroidal magnetic field that can
be supported by the plasma in the thin disk. Later, \citet{wang1995a}
derived three models based on different field configurations that are
consistent with the plasma conditions of the thin disk. Assuming that
the field is determined by re-connection at the Alfv\'en speed yields
\citep[][Eq.~9]{wang1995a}
\begin{equation}
n_{\tn{W}_\tn{I}}(\omega_{\tn{W}_\tn{I}}) = 1 +
\frac{1}{3}\frac{\omega_{\tn{W}_\tn{I}}^{57/40}}{\sqrt{1-\omega_{\tn{W}_\tn{I}}}}
\left(\int_{\omega_{\tn{W}_\tn{I}}}^1\frac{\sqrt{1-y}}{y^{97/40}}\tn{d}y
-\int_1^\infty\frac{\sqrt{y-1}}{y^{97/40}}\tn{d}y\right)\,.
\end{equation}
Assuming that the growth of the magnetic field in the disk is limited
by turbulent mixing gives \citep[see their Eq.~15]{wang1995a}
\begin{equation}
n_{\tn{W}_\tn{II}}(\omega_{\tn{W}_\tn{II}}) = 
1 + \frac{1}{6}\frac{1-2\omega_{\tn{W}_\tn{II}}}{1-\omega_{\tn{W}_\tn{II}}}\,,
\end{equation}
a result that was also obtained by \citet{yi1995a} to explain the spin
of T\,Tauri stars.
Finally, the case that the magnetic field re-connection happens
outside of the disk gives \citep[][their Eq.~19]{wang1995a}
\begin{equation}
n_{\tn{W}_\tn{III}}(\omega_{\tn{W}_\tn{III}}) =
1 + \frac{1}{18}\frac{2\omega_{\tn{W}_\tn{III}}^2-6\omega_{\tn{W}_\tn{III}}+3}{1-\omega_{\tn{W}_\tn{III}}}\,.
\end{equation}
The response field in the disk is assumed to be the same in all three
cases, while the timescale at which the internal field can react to
external changes is derived for different characteristic velocities of
the plasma (e.g., speed of sound).

For the configuration given by \citet{wang1995a} together with the
three characteristic velocities one finds the following implicit relations between $\hat\omega$ and $\omega_i$:
\begin{equation}
\begin{split}
\frac{\xi}{8} \omega_{\tn{W}_\tn{I}}^{14/3} + \gamma \eta^4
\sqrt{\frac{P_\tn{z}}{2}} \hat\omega^{14/3}
(\omega_{\tn{W}_\tn{I}} - 1) = 0 \\
\frac{\alpha}{\sqrt{8}}\omega_{\tn{W}_\tn{II}}^{7/3} +
\gamma \eta^2\hat\omega^{7/3} (\omega_{\tn{W}_\tn{II}} - 1)
= 0 \\
\frac{1}{\sqrt{8}}\omega_{\tn{W}_\tn{III}}^{7/3} +
\gamma_\tn{max} \eta^2 \hat\omega^{7/3}
(\omega_{\tn{W}_\tn{III}} - 1) = 0\,.
\end{split}
\end{equation}
Here all symbols are matched to the original description of
\citet{wang1995a}, except for the variable $P_\tn{z}$, that characterizes the
pressure of the plasma at the truncation radius in units of the
magnetic pressure perpendicular to the disk. Here, $\xi \leq 1$ is a
fraction of the local Alfv\'en speed, $\gamma \gtrsim 1$ characterizes
how quickly the local velocity transitions from Keplerian rotation to
co-rotation when leaving the disk plane, $\eta \leq 1$ is a screening
coefficient of the field in the disk, $\alpha \leq 1$ is similarly a
scaling factor for the local sound speed, and $\gamma_\tn{max}$ is the
maximum pitch value that can be maintained.

Since the models of \citet{wang1995a} and \citet{ghosh1979b} have a
discontinuity at $\omega_i = 1$, they do not allow for a transition to
the propeller state. \citet{rappaport2004a} argued that when $R_\tn{t}
\gtrsim R_\tn{co}$ the disk is not truncated further away, but instead
the inner edge will be located at $R_\tn{co}$. This argument is
supported by numerical simulations \citep{romanova2008a,ireland2022a}
and in line with the arguments of \citet{perna2006a}. Based on this
assumption, \citeauthor{rappaport2004a} constructed a model that
extends the configuration described by \citet{wang1995a}. Effectively
the model limits $\omega_i \leq 1$ with the extension that for $R_\tn{t}
> R_\tn{co}$ $\omega_i = 1$, such that \citet[their Eq.~23 \&
  24]{rappaport2004a} is rewritten as
\begin{equation}
n_\tn{R}(\omega) = 1 +
\frac{1}{9}\left(2\omega_\tn{R}^2-6\omega_\tn{R}+3\right)\,,
\end{equation}
where
\begin{equation}
\omega_\tn{R} = 
\begin{cases}
(1/2)^{3/14}\hat\omega\, & \mbox{for $\hat\omega \leq 1$} \\
1 & \mbox{otherwise.}
\end{cases}
\end{equation}
It is important to note that for all models considered so far, a
discontinuity is only reached if $\omega_i$ can equal unity. Even the
model of \citet{wang1995a} has the property that $\omega \rightarrow
1$ only for $\dot{M} \rightarrow 0$. Similarly, the full derivation of
\citet{ghosh1979b} only approaches unity asymptotically. However, due
to the approximations applied by \citeauthor{ghosh1979b}, which
allowed them to derive analytic expressions, it appears as if $\omega_i$
is unrestricted. As a results, the spin-down
torque for large spin frequencies or low mass-accretion rates will be overestimated because $\omega$ should approach $1$ only asymptotically. As discussed by \citet{rappaport2004a}, their definition of $\omega_i$ is based on setting $\Gamma = (1/2)^{1/7}$ in
Eq.~(\ref{eq:rt-linear-fix}). Note that the definition of the Alfv\'en radius by \citeauthor{rappaport2004a} misses a factor of two in the denominator with respect to our definition in Eq.~(\ref{eq:ralfven}).

Another model using exactly the same concepts as the ones discussed up
to this point was introduced for T\,Tauri stars \citep{matt2005a},
where the magnetic field of a young star limits its maximum rotation
rate. While the conditions in these systems are generally different
from those in accreting neutron star systems, the accretion process is
described by the same ideas. The model of \citet{matt2005a} is
parameterized in a way that determines the coupling strength between
the disk and the magnetic field with two parameters. The inner
connected region is determined by the maximum pitch angle,
$\gamma_\tn{c}$, that can be maintained for the lines connecting to
the disk. A second parameter characterizes the diffusion of the
plasma, $\beta$, and effectively models the coupling strength between
the disk and the central field. Large values of $\gamma_\tn{c}$ permit
more extended regions of the disk to be connected. From this,
\citet[Eq.~19 \& 21]{matt2005a} derived a parametric torque function
\begin{equation}
  n_\tn{M}(\omega_\tn{M}) =
\begin{cases}
  1 +
  \frac{2}{3}\frac{2\omega_\tn{M}^2\omega_\tn{p}^{-1}-\omega_\tn{M}^2\omega_\tn{p}^{-2}-2\omega_\tn{M}+1}{1-\omega_\tn{M}}
  & \mbox{for $\omega_\tn{M} \geq \omega_\tn{c} = 1-\beta\gamma_\tn{c}$} \\
  n_\tn{M}(\omega_\tn{c}) & \mbox{elsewhere,} \\
\end{cases}
\end{equation}
where $\omega_\tn{p} = 1+\beta\gamma_\tn{c}$. Due to the
chosen parametrization, this model is able to reproduce most other models
described here.

The defining equation for the fastness in the model of \citet{matt2005a} can be written as
\begin{equation}
\begin{split}
\omega_\tn{M} = 1 - \beta\gamma & \quad\tn{if } \hat\omega <
(\sqrt{8}\gamma)^{-3/7}(1-\beta\gamma) \\
\frac{\beta}{\sqrt{8}}\omega_\tn{M}^{7/3} +
\hat\omega^{7/3}(\omega_\tn{M} - 1) = 0 & \quad\text{otherwise}\,,
\end{split}
\end{equation}
where $\gamma$ is again the magnetic pitch angle and $\beta$
characterizes the coupling between the magnetic field and
the disk, where small values indicate strong coupling.

By assuming that the magnetic configuration of \citet{wang1995a} holds
for all cases and that the torque is given from the Kepler orbits
beyond $R_\tn{co}$, \citet{dai2006a} made another attempt to allow
transitions to $\omega_i > 1$. For an arbitrarily chosen exponential
transition function, they find \citep[their Eq.~11]{dai2006a}
\begin{equation}
n_\tn{D}(\omega_\tn{D}) = 
\left\{\begin{array}{lr}
\xi(1-\omega_\tn{D})+\frac{\sqrt{2}\gamma}{3}
\left(\frac{2}{3}\omega_\tn{D}^2-2\omega_\tn{D}+1\right) & \mbox{if $\omega_\tn{D}
\leq 1$} \\
\xi(1-\omega_\tn{D})+\frac{\sqrt{2}\gamma}{3}
\left(\frac{2}{3}\omega_\tn{D}-1\right) & \mbox{otherwise}\,.
\end{array}\right.
\end{equation}
where $\xi \lesssim 1$ is a structure factor defining the torque at
$R_\tn{t}$ and where $\gamma$ is the limit of the magnetic field pitch
at $R_\tn{t}$. This is yet another model using the linear approximation
(Eq.~\ref{eq:rt-linear-fix}), but this time with $\Gamma = 1$, that is,
\begin{equation}
\omega_\tn{D}=\hat\omega
\end{equation}
Combining the arguments of \citet{rappaport2004a} with the description
of \citet{wang1995a}, \citet[][their Eq.~36]{kluzniak2007a} derived a
torque function very similar to $n_{\tn{W}_\tn{III}}$,
\begin{equation}
n_{\tn{K}_\tn{I}}(\omega_{\tn{K}_\tn{I}}) = 1 +
\frac{1}{18}\frac{2-3\omega_{\tn{K}_\tn{I}}}{1-\omega_{\tn{K}_\tn{I}}}\,.
\end{equation}
Importantly, based on the $\alpha$-disk model, \citeauthor{kluzniak2007a} solved the disk structure
for their assumptions as well as for the magnetic field configuration
of \citet{wang1995a}. They also derived
the torque based on the same assumptions as \citet{wang1995a},
therefore their second model, $n_{\tn{K}_\tn{II}}$, is equal to
$n_{\tn{W}_\tn{III}}$. The defining equations for the fastness are given by
\begin{equation}
\begin{split}
\frac{1}{\sqrt{8}}\omega_{\tn{K}_\tn{I}}^{10/3} +
\hat\omega^{7/3}(\omega_{\tn{K}_\tn{I}} - 1) = 0 \\
\frac{1}{\sqrt{8}}\omega_{\tn{K}_\tn{II}}^{7/3} +
\hat\omega^{7/3}(\omega_{\tn{K}_\tn{II}} - 1) = 0\,.
\end{split}
\end{equation}
Concentrating on the strong magnetic fields present in ultra-luminous
X-ray sources, \citet{gao2021a} extended the arguments of
\citet{wang1987a,wang1995a}. Instead of assuming Keplerian motion at
$R_\tn{t}$, they propose a stiff disk model where the velocity at
$R_\tn{t}$ matches the rotation velocity of the compact object
(instead of the local Keplerian velocity). Otherwise following the
same arguments of \citet[for two of the three characteristic velocities
  given there]{wang1995a} they found \citep[Eq.~10 \& 12]{gao2021a}
\begin{equation}
\begin{split}
n_{\tn{G}_\tn{I}}(\omega_{\tn{G}_\tn{I}}) &= 1 +
\frac{2}{3}\frac{1-2\omega_{\tn{G}_\tn{I}}}{1-\omega_{\tn{G}_\tn{I}}} \\
n_{\tn{G}_\tn{II}}(\omega_{\tn{G}_\tn{II}}) &= 1 +
\frac{2}{9}\frac{2\omega_{\tn{G}_\tn{II}}^2-6\omega_{\tn{G}_\tn{II}}+3}{1-\omega_{\tn{G}_\tn{II}}}\,.
\end{split}
\end{equation}
The fastness is implicitly given via
\begin{equation}
\begin{split}
2\alpha \omega_{\tn{G}_\tn{I}}^{7/3} +
\gamma\eta^2\hat\omega^{7/3} (\omega_{\tn{G}_\tn{I}} - 1) = 0 \\
2 \omega_{\tn{G}_\tn{II}}^{7/3} +
\gamma\eta^2\hat\omega^{7/3} (\omega_{\tn{G}_\tn{II}} - 1) = 0\,, \\
\end{split}
\end{equation}
where the parameters have the same meaning as for the model
of \citet{wang1995a}.

\subsection{Model comparison}

\begin{figure}
\resizebox{\hsize}{!}{\includegraphics{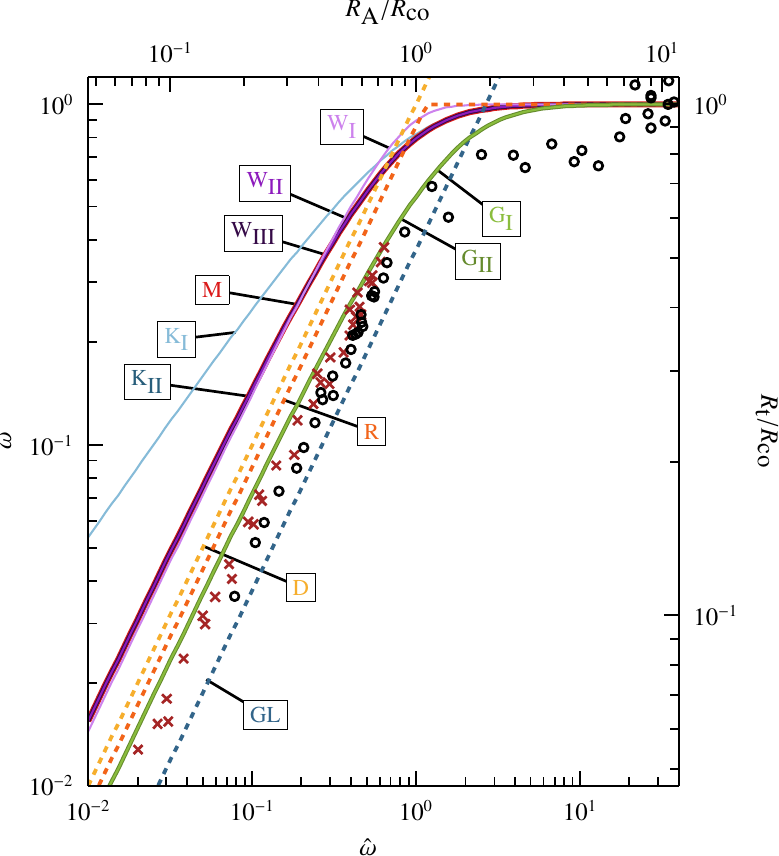}}
 \caption{Relation between the model-dependent fastness, $\omega_i$, and the canonical
 fastness, $\hat\omega$, as predicted or assumed for the different models.
 Labels indicate the model with corresponding subscript. Dashed
 lines are models that assume $R_\tn{t} \propto R_\tn{A}$.
 Red crosses indicate the values from the simulations of
 \citet{ireland2020a}, black circles are from
 \citet{ireland2022a}.}
\label{fig:fastness-default}
\end{figure}

Having expressed the torquing models in a uniform manner, we can now perform a comparison between them that is based on an identical meaning of the dynamical variable governing the torque, $\hat\omega$. The relation between the model-dependent $\omega_i$ and $\hat\omega$ for all considered
models is shown in Fig.~\ref{fig:fastness-default}. For models with additional parameters, we chose the default parameters by setting the additional model parameters to unity, with the exception of the model
of \citet{matt2005a} where $\beta = 1$ and
$\gamma_\tn{c} \rightarrow \infty$ (causing $R_\tn{out} \rightarrow \infty$).
The figure also includes results from \citet{ireland2020a,ireland2022a} that
have been obtained from numerical simulations of accreting T\,Tauri stars.
In the T\,Tauri star community the disk-field interaction is invoked to explain
the observed limit of the rotational period of the accretors that is much
lower than what would theoretically be possible when considering only
the amount of accreted matter with angular momentum given by Keplerian motion.
The idea is that the connection of the central field to the disk causes
currents in the disk that, in turn, create a magnetic field
\citep{goodson1999a, goodson1999b}. At the boundary layer the magnetic
field reconnects, trapping plasma from the disk. When the magnetic field
is unloaded to release the stress it causes magnetospheric ejections
\citep{zanni2013a}. These magnetospheric ejections carry away large amounts
of angular momentum and, thus, limit the maximum spin of the accretor.

Except for the models shown with dashed lines, and from
\citeauthor{kluzniak2007a}, there are usually two to three parameters
for each model that influence the detailed behavior and can
essentially be referred back to the plasma parameters in the disk.
Irrespective of the detailed behavior the global trend of $\omega_i(\hat\omega)$ is always the
same: For $\hat\omega \ll 1$ the fastness is a linear function, roughly
matching the estimate from \citet{ghosh1979b}, but often with an offset. For larger values, that
is, lower mass accretion rates for a given spin frequency, the
dependency changes and $\omega_i$ approaches unity asymptotically. This fact
seems to be often overlooked in the literature, given the number of
models which extrapolate the linear relation to fast systems. Rather, for $\hat\omega\rightarrow\infty$ most models will lead to the truncation radius approaching $R_\tn{co}$, matching the argument of
\citet{rappaport2004a}.

\begin{figure}
\resizebox{\hsize}{!}{\includegraphics{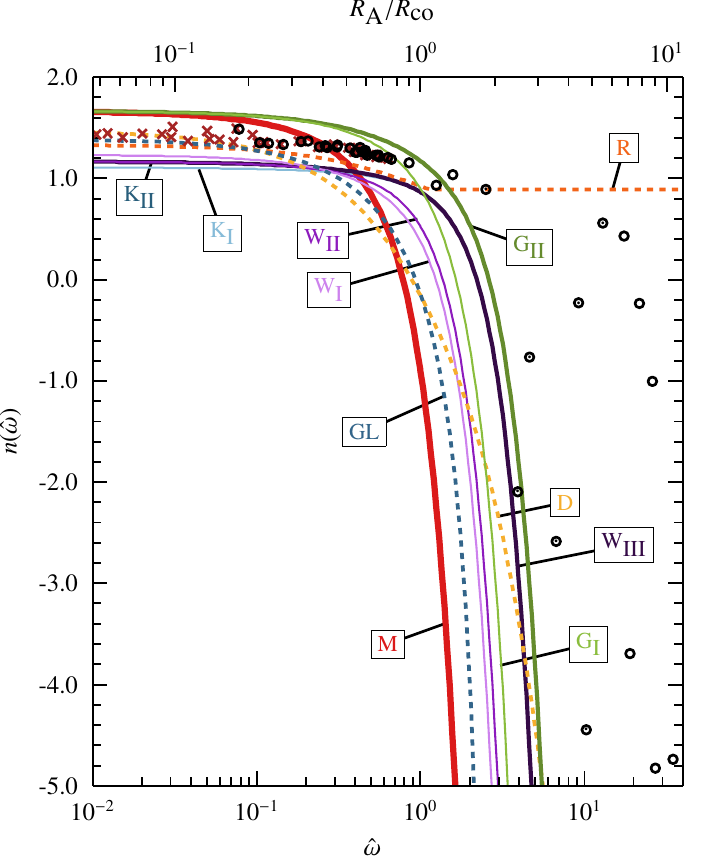}}
\caption{Torque function of all considered models expressed via the
  canonical fastness. Labels indicate the models with the
  corresponding subscript. Dashed lines indicate models where
  $R_\tn{t} \propto R_\tn{A}$. Red crosses mark the results of
  \citet{ireland2020a} and black circles that of \citet{ireland2022a}.
  The black circles with a dot identify the simulations in the
  transition region of Fig.~\ref{fig:fastness-default}, with $0.6 <
  \omega_i < 0.8$.}
\label{fig:torque-canonical-default}
\end{figure}

The resulting torque functions expressed as a function of the canonical fastness
are shown in Fig.~\ref{fig:torque-canonical-default}. The figure illustrates that, with the exception of the model of \citet{rappaport2004a}, the different models are far more similar in behavior than one might have inferred from Fig.~\ref{fig:torque-default}.  Two quantities can be readily extracted from the models that show
the remaining differences between them especially well: the maximum torque value possible and the
equilibrium point, where $n(\omega_i) = 0$. Compared with the simulations of
\citet{ireland2020a,ireland2022a} it seems that most models drop off
too quickly. This might be possible to resolve by employing different
parameters for the magnetic interaction for the different models. What is
striking is that the points of the simulation that are close to or
beyond the equilibrium condition correspond to the points where the
fastness deviates from the model predictions (dots in
Fig.~\ref{fig:torque-canonical-default}). It seems that the analytical
models are no longer valid close and beyond the equilibrium point,
that is, for spin down. However, as said before, it is not clear
whether the conditions of T\,Tauri stars are comparable to those
in compact objects. Furthermore, the resulting torque from the
simulations depends on how it is extracted from the results. Given the
chaotic behavior at spin down, this might introduce additional biases.

It is also noteworthy that the
simulations have a maximum torque value at $\omega_i \rightarrow 0$ that
exceeds the analytical models, except for the expressions of
\citet{matt2005a} and \citet{gao2021a}. The reason is the assumed velocity
at $R_\tn{t}$. In most models it is assumed that up to this radius,
the disk is described by Keplerian motion, while for $n_\tn{G}$ it is
assumed that the velocity at the inner edge of the disk matches that
of the rotating magnetic field. A special case is the model of
\citet{matt2005a}, where the coupling between the disk and the field
are subject to the value of $\omega_\tn{p}$. This behavior also
influences how strongly the plasma is forced to follow the field. For
smaller values the disk experiences more drag, specifically for
$\omega_\tn{p} = 2$, $n_\tn{M} = n_{\tn{G}_\tn{I}}$. The other
solution, $n_{\tn{G}_\tn{II}}$, is approximately reproduced by
$n_\tn{M}$ for $\omega_\tn{p} = 1+ \sqrt{3/2}$. Due to the branch for
the critical fastness, $\omega_\tn{c}$, a somewhat similar behavior to
the \citeauthor{wang1995a} models can be achieved, however, with a
constant spin-up below the critical fastness and a sharper drop off
above.

\subsection{Torquing neutron stars: Predicting spin changes}

Having established the different values for the torque, we can now compute how the accretion flow changes the spin of the neutron star. 
The torque relation (Eq.~\ref{eq:spin-up}) connects the spin period
and its derivative with the system parameters (magnetic field,
mass, distance, and spin period) and the mass accretion rate. As the
latter cannot be observed directly, it is commonly assumed that it
correlates with the X-ray luminosity of accreting systems
\citep{zeldovich1966a,shklovsky1967a}.
Assuming that the potential energy of the in-falling matter is
converted to photons with X-ray luminosity
\begin{equation}
\label{eq:xlum}
L_\tn{X} = \varepsilon\frac{\dot{M} G M}{R}\,,
\end{equation}
where $\varepsilon \leq 1$ is the conversion efficiency\footnote{The
conversion efficiency, $\varepsilon$, is commonly set to 1 for
accreting neutron stars.}, $R$ the radius of the accretor, and $M$ its
mass. The luminosity is obtained from the observed flux, $F$, by
\begin{equation}
\label{eq:fluxtolum}
L_\tn{X} = \alpha F D^2\,,
\end{equation}
where $D$ is the source distance and $\alpha$ is a parameter depending
on the emission characteristic. For isotropic emission
$\alpha = 4\pi$.
Using Eq.~(\ref{eq:keplertorque}), (\ref{eq:xlum}), and
(\ref{eq:fluxtolum}), and expressing the moment of inertia via the
dimensionless radius of gyration, $k$,
\begin{equation}
\label{eq:rgyration}
I = (k R)^2 M,
\end{equation}
we obtain
\begin{equation}
-2\pi k^2 \sqrt{G} h \dot{P} = \frac{P^2}{R M^{3/2}} D^2 F R_\tn{t}^{1/2}
\end{equation}
where $h = \varepsilon/\alpha$, while
the definition of the fastness and the canonical fastness yield
\begin{equation}
R_\tn{t} = \left(\frac{\omega}{\hat\omega}\right)^{2/3} R_\tn{A}\,.
\end{equation}
Combining this equation and Eq.~(\ref{eq:ralfven}) we can finally write
\begin{equation}
\label{eq:spin-up-fit}
-2\pi k^2 \left(G^3\sqrt{2}\right)^{1/7} \frac{\tn{d}P}{\tn{d}t} = 
\left(\frac{\omega}{\hat\omega}\right)^{1/3}
\left(\frac{\mathcal{A}}{\mathcal{B}}\right)^2 n(\omega)
\left(P F^{3/7}\right)^2
\end{equation}
where we introduced two auxiliary parameters, the \emph{effective pole strength},
$\mathcal{A}$, 
\begin{equation}
\mathcal{A} = \frac{\mu}{MR} 
\sim 10^{24}\,\tn{G\,cm}^2\,M_\odot^{-1} \left(\frac{\mu}{10^{30}\,\tn{G}\,\tn{cm}^3} \right) \left(\frac{M}{M_\odot} \right)^{-1} \left(\frac{R}{10^6\,\tn{cm}} \right)^{-1}
\end{equation}
and the \emph{coupling strength}, $\mathcal{B}$,
\begin{multline}
\mathcal{B} =
\left(\frac{\mu^6}{R^3M^2(hD^2)^3}\right)^{1/7}\\
\sim 3.59 \times 10^{37}\,\tn{kpc}^2\,\tn{g}^2\,\tn{cm}^{-1}  h^{-3/7}\left(\frac{\mu}{10^{30}\,\tn{G\,cm}^3}\right)^{6/7} \left(\frac{R}{10^6\,\tn{cm}} \right)^{-3/7}\\
\left(\frac{M}{M_\odot} \right)^{-2/7} \left(\frac{D}{1\,\tn{kpc}} \right)^{-6/7}\,,
\end{multline}
which summarize the system parameters

Combining the conversion factor, $\varepsilon$, and the emission
characteristic, $\alpha$, into the \emph{weighted efficiency}, $h$,
expresses our ignorance towards the relation between accreted matter
and photons reaching the observer, including the limited energy range
of the instruments. Generally $h$ is time dependent, but often assumed
to be constant for simplicity. We expect that deviations from constancy are
mainly influenced by long term trends in the emission characteristic,
that is, changes in $\alpha$.  The shift of maximum emission relative to
the energy band used to measure $L$ is reflected by changes in
$\varepsilon$. Finally, an effect that seems often ignored when torque models
are applied to data obtained from neutron stars is that when the
mass-accretion rate is calculated based on the observed flux, the
gravitational red-shift must be taken into account, since the flux
observed differs from the emitted flux in the frame of rest of the
neutron star by a factor between 1.6 and 2.4. In our notation this
correction is implicitly taken into account through the value of $h$.

The particular separation into the two auxiliary variables is done such
that the canonical fastness is only dependent on one of those
parameters
\begin{equation}
\hat\omega =
2\pi\left(\sqrt{2}^3G^2\right)^{-1/7}\frac{\mathcal{B}}{PF^{3/7}}\,.
\end{equation}
In this form the model can be directly compared with data
consisting of the time-dependent flux, $F$, and spin period,
$P$. The differential equation~(\ref{eq:spin-up-fit}) can be solved by standard
numerical integration. With a particular choice of the torque
function $n(\omega)$ and corresponding relation between
$\hat\omega$ and $\omega$ the model can be fitted to the data.

\subsection{Equilibrium fastness}
\label{sec:equilibrium}

\begin{figure}
\resizebox{\hsize}{!}{\includegraphics{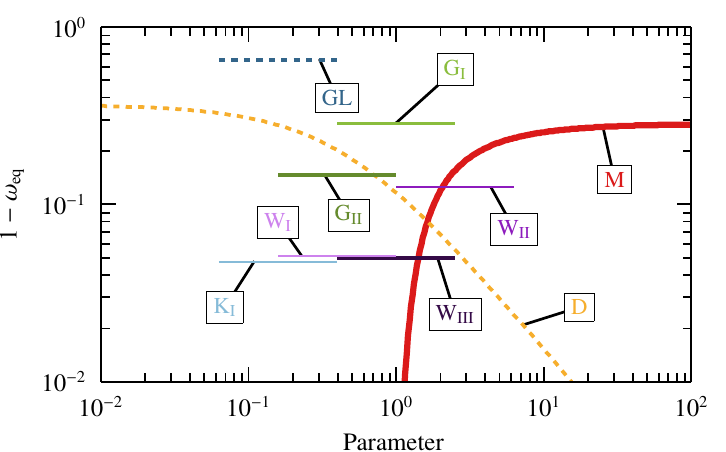}}
\caption{Visual representation of the equilibrium fastness.
The ``parameter'' is either $\omega_\tn{p}$ for
\citet{matt2005a}, or $\xi/\gamma_\tn{c}$ for
\citet{dai2006a}. For $\omega_\tn{p}$ the lower bound is 1.
All other models have fixed values for the equilibrium fastness
(see Table~\ref{tab:equilibrium}) and
are displayed over an arbitrary short range for clarity only. Dashed lines
indicate models where $R_\tn{t} \propto R_\tn{A}$.}
\label{fig:equilibrium}
\end{figure}

Given all models above we can immediately calculate
the equilibrium point, $\omega_\tn{eq}$, where
$n(\omega_\tn{eq}) = 0$. If a source is accreting constantly
the equilibrium value corresponds to a spin period
$P_\tn{eq}$ that should be reached after some time, and here
the central accreting object is neither spinning up nor
down. This value is often used to estimate system parameters
as it is expected that persistent sources accrete roughly
with a constant rate. For this reason we give the
equilibrium fastness for all models in
Table~\ref{tab:equilibrium}. To visualize how the
equilibrium location changes for the parameters for the
models of \citet{matt2005a} and \citet{dai2006a} with
respect to the other models, Fig.~\ref{fig:equilibrium}
shows the values for different parameter combinations, where
the $x$-axis is $\omega_\tn{p}$ for $n_\tn{M}$ and
$\xi/\gamma_\tn{c}$ for $n_\tn{D}$. As for the torque function,
comparison between different models for a particular system must
be done via $\hat\omega$.

\begin{table}
\caption{Resulting equilibrium fastness values,
$\omega_\tn{eq}$.}
\label{tab:equilibrium}%
\def\arraystretch{1.3}%
\begin{tabular*}{0.7\hsize}{@{\extracolsep{\fill}}rl}
\hline
Model & $\omega_\tn{eq}$ \\
\hline
GL & 0.349\tablefootmark{a)} \\
W$_\tn{I}$ & 0.949\tablefootmark{a)} \\
W$_\tn{II}$ & $7/8$ \\
W$_\tn{III}$ & $6-\sqrt{3\cdot17/2} \approx
0.950$ \\
R & N/A\tablefootmark{b)} \\
M & \(\displaystyle\left\{\begin{array}{l}
 \frac{7\omega_\tn{p}^2 -
 \omega_\tn{p}\sqrt{(7\omega_\tn{p})^2-40(2\omega_\tn{p}-1)}}{4(2\omega_\tn{p}-1)}
 \\
 5/7 \tn{ for } \gamma_\tn{c}\rightarrow\infty
 \end{array}\right.\)
\\
D & $1+c-\sqrt{1/2+c^2}$, $c =
\frac{9}{4\sqrt{2}}\frac{\xi}{\gamma_\tn{c}} + 1/2$ \\
K$_\tn{I}$ & $20/21$ \\
G$_\tn{I}$ & $5/7$ \\
G$_\tn{II}$ & $\frac{21 - \sqrt{3\cdot67}}{8} \approx 0.853$ \\
\hline
\end{tabular*}

\tablefoottext{a}{Equilibrium value is calculated numerically
via root finding.}

\tablefoottext{b}{This model never crosses over to spin-down.}
\end{table}

\begin{figure}
\resizebox{\hsize}{!}{\includegraphics{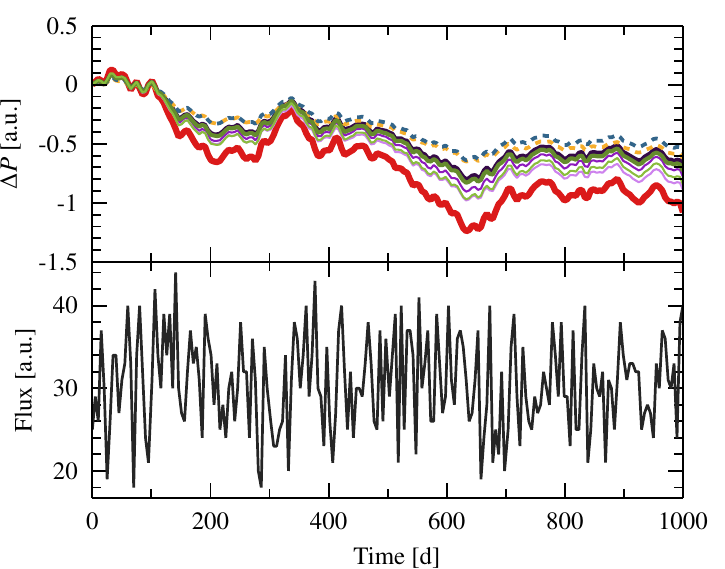}}
\caption{Top panel: Predicted change in rotation period for
all models \citep[except][]{rappaport2004a} for a given starting period and a mass accretion rate defined by a Poisson process. Colors indicate
the models, as before. Dashed lines indicate models with
$R_\tn{t} \propto R_\tn{A}$. Bottom panel: Simulated Poisson
light-curve.} \label{fig:equilibrium-period}
\end{figure}

One important factor of the equilibrium fastness is to
understand how it connects to a potential equilibrium
period. The latter is often used as an estimate for
persistently accreting systems. However, the equilibrium
fastness only directly translates into an equilibrium period
if the mass accretion rate is constant. More generally, however, the
mass accretion rate can often be approximated by a stochastic process, potentially with a
constant mean accretion rate. Depending on the exact parameters of
the system, this stochastic process can lead to a drift of the
period instead of to a system in equilibrium, even if the average mass accretion rate is constant. As a simple
demonstration, Fig.~\ref{fig:equilibrium-period} shows how the period of a neutron star is predicted to change with time for the different torquing models. Despite the fact that the average mass accretion rate is constant in the Poisson process assumed here, and despite the fact that the starting period was set to match the equilibrium period for that mass accretion rate, the period of the neutron star is predicted to vary with time. 

\section{Application to the Be X-ray binary \texorpdfstring{4U\,0115$+$63}{4U 0115+63}}
\label{sec:example}

We now apply torquing theory to a real observational example, in order to illustrate how they can be used to estimate system parameters and in order to reveal some of the typical problems that can occur using torque models.
Specifically, we consider the four observed outbursts of the Be binary system
\object{4U\,0115$+$63} \citep[and references therein]{giacconi1972a,kuehnel2020a}
captured by
Fermi/GBM\footnote{\url{https://gammaray.msfc.nasa.gov/gbm/science/pulsars/lightcurves/4u0115.html}.}
\citep{meegan2009a} and
Swift/BAT\footnote{\url{https://swift.gsfc.nasa.gov/results/transients/weak/4U0115p634/}./}
\citep{gehrels2004a,gehrels2005a}. This source is very well known
because of its multiple CRSFs in the X-ray spectra that allow for
accurate measurement of the magnetic field strength \citep{wheaton1979a,santangelo1999a}.
The discussion here is only to showcase the performance of the models
in comparison to each other and should not be understood as an
in-depth analysis of the system. Besides the direct measurement of the
magnetic field in 4U\,0115 through cyclotron line measurements, being a
Be-binary system with several bright outbursts, the source shows a very clear spin-up
signal linked to the large dynamic range in mass accretion rate. In comparison,
more classically viewed disk accretion systems like Vela~X-1 or Cen~X-3 are close
to the spin equilibrium and the accretion is therefore more chaotic causing
some disconnection between the spin and luminosity signal. As we illustrated 
in Sect.~\ref{sec:torque-models} it should be clear that the prediction for spin-down
is much less understood which would complicate the comparison even more.

Before attempting to combine the light-curve, the observed evolution
of the spin period, and a specific torque model, we can perform
expectation management based on the expression given in
Eq.~(\ref{eq:spin-up-fit}). Generally, for accreting highly-magnetized
neutron stars, we can assume that the mass, radius, moment of
inertia, and magnetic field of the accreting object are constant in time
due to the overall small accreted mass and the time-scales considered
here. With this in mind, the only remaining time
dependence in the parameters is the conversion efficiency and the
emission characteristic in $\mathcal{B}$. Without further information
on the flux, we assume that $h$ is constant on timescales much larger
than the pulse period. This leaves us with a total of five parameters, one $\mathcal{A}$, and four $\mathcal{B}$'s.

\subsection{Confronting reality}

In order to obtain the mass accretion rate from the observed
light-curve, a conversion of the instrument specific count rate to
luminosity is necessary. If broad band spectral coverage is available,
the luminosity can be determined from the measured flux by assuming a
certain emission characteristic. Here we will simply assume that this
conversion factor is constant over each outburst, it
can therefore be seen as an additional contribution to $h$. For
an unconstrained parameter search, we make no assumptions on the
distance and determine the value of $\mathcal{B}$ directly from the
data.

Equation~(\ref{eq:spin-up-fit}) directly shows that there is a
parameter correlation between $\mathcal{A}$ and $\mathcal{B}$ when model fitting. Given that for
slow rotators $n$ tends towards a constant value, it is nearly
impossible to disentangle the two parameters without additional
information. Strong correlation between parameters poses a challenging
problem for most fit methods. For this reason it is better to treat
the ratio $\mathcal{A}/\mathcal{B}$ as one parameter. But even with
this modification a correlation between the parameters remains, since
for larger values of $\mathcal{B}$ the canonical fastness will increase. This,
in turn, reduces the spin-up and increases $\mathcal{A}/\mathcal{B}$.
Only the transition to spin-down at even larger values of
$\mathcal{B}$ limits the range of the parameters. Overall, it should
therefore be expected that a large systematic uncertainty will be
present in all fit results, introduced by the conversion from
light-curve to mass accretion rate. An unconstrained parameter search
will thus most likely never result in specific estimates for the magnetic
field or distance but only a relation between those two quantities.

Another factor complicates the matter: Binary motion. Be X-ray binaries
have typical orbital periods of 10--1000\,days \citep{malacaria2020a}.
In some systems regular type I outbursts can be observed tightly linked
to the orbital period, that last only a fraction of the orbit. More
luminous type II outbursts are observed rarely and less connected with
the orbital period, and can last over serval period cycles \citep{reig2011a}.
Especially for the latter the measured spin period is affected by the Doppler
shift due to the orbital motion. To predict the change in spin period the
mass accretion rate at the rest frame of the accretor must be
determined. To convert the flux from the observer frame to the rest
frame the a solution for the orbit must be found. Unless the orbit is
well known from other measurements, the best way to achieve this is by
fitting the orbital parameters simultaneously to the torque function.
The full model is then written as
\begin{equation}
P_\text{obs}(t_\text{obs}) = D(t_\text{rest})
(T(t_\text{rest},F(t_\text{rest})) + P_0)\,,
\end{equation}
where $T$ is the solution of Eq.~(\ref{eq:spin-up-fit}) with
integration constant $P_0$, $D$ is the Doppler correction of
the intrinsic period due to orbital motion, and the
connection between $t_\text{obs}$ and $t_\text{rest}$ is
obtained from the light travel time of the orbit solution. For
many objects the orbital parameters are not well determined
to allow a correction upfront. Therefore applying the orbit as
part of the model is necessary and also allows to measure a
potential time dependence of the orbital elements.

\subsection{Torquing results}

\begin{figure*}
\includegraphics[width=\hsize]{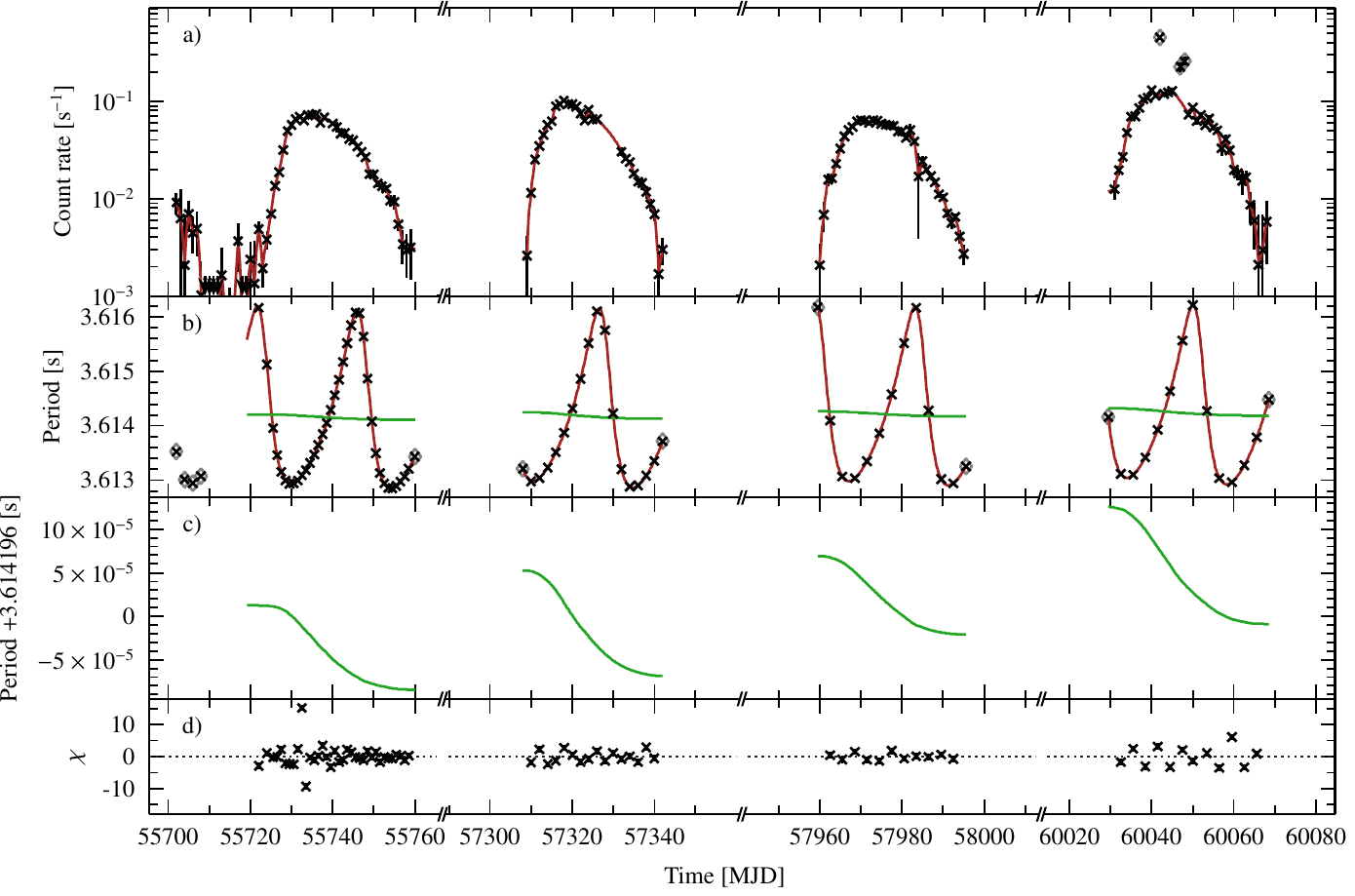} \caption{Torque
  modeling for 4U\,0115+63. (a) Light-curve of 4U\,0115$+$63
  as measured by Swift/BAT. The red curve is interpolation of the
  data. (b) Pulsation period as measured by Fermi/GBM. The
  red curve is the best fit model of W$_\tn{III}$ modified by the
  Doppler shift of the orbit. The green curve is the intrinsic spin-up
  of the NS. (c) The intrinsic spin up signal predicted by
  the model. Same as the green curve in the panel above. d)
  $\chi$ residuals of the spin model. For all panels, the crosses
  enclosed by gray diamonds have been ignored for the model comparison
  and also note the gaps on the time axis.}
\label{fig:4u0115-lc}
\end{figure*}

We show the light-curve of 4U\,0115$+$63 as monitored by Swift/BAT in
Fig.~\ref{fig:4u0115-lc}, together with the spin period as measured by
Fermi/GBM. Only the relevant time spans of bright phases are shown.
\citet{wang1995a} argue that their model III should describe the
scenario of an accreting Be X-ray binary. The resulting best fit of
this model modified by the Doppler effect is shown in red, the
intrinsic spin-up is shown in green. We ignored several data points
from the Fermi/GBM data, mainly because they are at the edge of the
intervals covered by Swift/BAT. The first few data points taken during
the outburst of 2011 are ignored because of the missing light-curve
information in between. We also ignored the three outliers in the
Swift/BAT data of the 2023 outburst.

Despite showing only the model of \citet{wang1995a}, we applied all
discussed models to the data in the same way. Without constraining the
magnetic field, all models fit the data very well, and to an extent
that no favorable model can be chosen.

To get a better understanding of the correlation between the
parameters we explore the statistics landscape of the parameter space
with the method proposed by \citet{foremanmackey2013a}. For all four
outbursts this reveals the already expected nearly linear correlation
between $\mathcal{A}$ and $\mathcal{B}$. Furthermore, the
orbital parameters show a correlation with the torque parameters.
While the resulting orbital parameters are hardly affected, it appears
that the torque parameters are more sensitive to the additional
degrees of freedom. To some extent this is expected as the impact of
the orbit on the signal is much larger than the impact of the spin up.
Some results of the parameters space exploration are shown in
Appendix~\ref{apx:mcmc}.

\begin{figure}
\resizebox{\hsize}{!}{\includegraphics{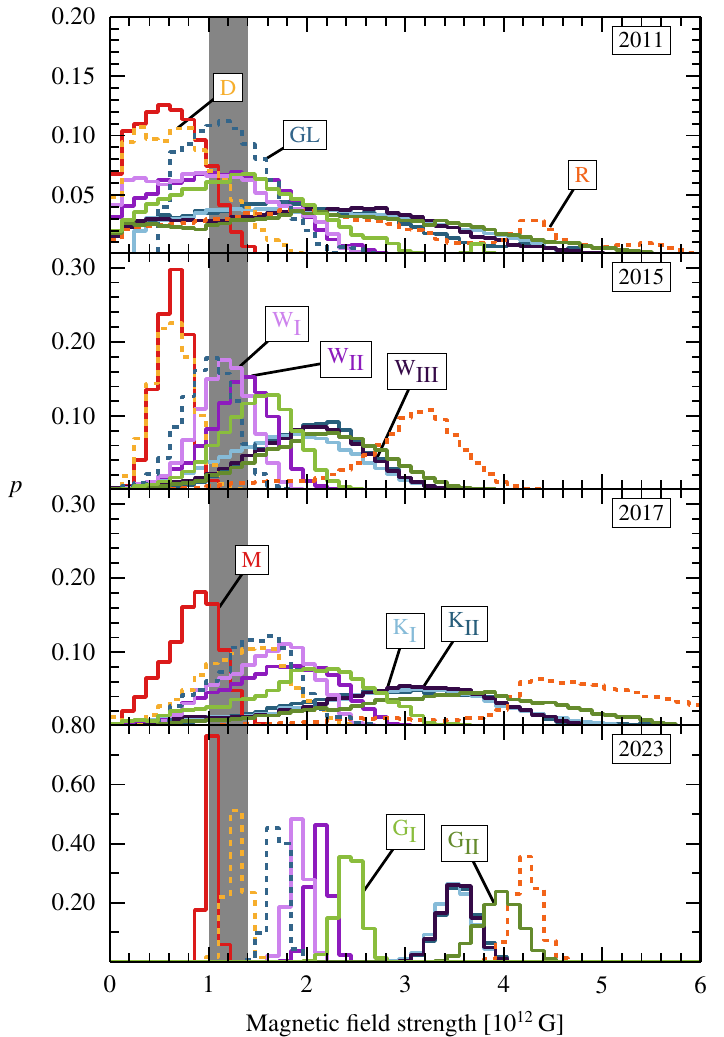}}
\caption{Marginal probability distribution of the pole
magnetic field strength estimated for all considered models.
The colors indicate the identifier of the model. Dashed
lines correspond to models with $R_\tn{t}\propto R_\tn{A}$. The gray
area indicates the value range estimated from the CRSF features.}
\label{fig:magnetic}
\end{figure}

From the distribution of the resulting torque parameters we can
estimate the distribution of the surface magnetic field at the pole
via $B = 2 M R^{-2} \mathcal{A}$. Assuming canonical values $M =
1.4\,M_\odot$ and $R = 10^6$\,cm we obtain the magnetic field
distributions for all models and outbursts (Fig.~\ref{fig:magnetic}). These distributions show clearly that any
magnetic field estimate that is based on torque theory is not well
constrained for a single model, and that magnetic field strength
estimates vary widely between different models. In addition, the
estimate suffers from a systematic uncertainty due to variations in
the plasma parameters. As such, it seems a far stretch to get reliable
values for the dipole moment. One might be tempted to argue that this
is only the case for our specific example, because no information of
the distance or the conversion factor is taken into account. However,
as is clear from the parameter dependencies, knowledge of this
information helps only if the emission characteristic and conversion
efficiency, $h$, are known too, which in general is not the case.

Torque models are often also used to measure the distance of a
source. Using parameter distributions we find that the distance
distribution between the outbursts is indeed very similar. However,
different models also give different estimates. Especially models
where $R_\tn{t} \propto R_\tn{A}$ consistently yield the largest
distances. This can be explained due to the fast transition to
spin-down present in these models. Generally these models require a
smaller $\mathcal{B}$ to describe the data, resulting in a larger
distance for the same $\mathcal{A}$.

The surface magnetic field of 4U\,0115$+$63, as estimated
from cyclotron line measurements \citep[for example]{kuehnel2020a} is
$1$ -- $1.4\times 10^{12}$\,G (depending on the assumed gravitational
redshift). In principle we can use these values to judge the performance of 
torque models to estimate magnetic field values.
From the distribution of the best-fit magnetic fields shown in
Fig.~\ref{fig:magnetic} it seems that the models group into three
classes. The lowest field estimates are generally given by the model
of \citet{matt2005a} and very similar by that of
\citet{dai2006a}. Slightly larger values are obtained with the
prescriptions of \citet{ghosh1979b} and the first and second model of
\citet{wang1995a}. The largest estimates, and also the least
constrained ones, are obtained based on \citet{kluzniak2007a} and
\citet{gao2021a}. The ad-hoc solution given by \citet{rappaport2004a}
shows very clearly the effect of the artificial fix point resulting
in a discontinuity in the torque function. In conclusion, while all models \citep[except][]{rappaport2004a} yield values in the ballpark of the cyclotron line results,  it seems that the third class of
models favors larger values. In comparison with the value obtained through cyclotron line measurements we find a first indication what torque models perform better. However, it seems that at least some models result in $B$-field estimates which are not compatible between the outbursts, which lets us conclude that those are affected by systematic uncertainties. Because it is not clear how
comparable the local magnetic field strength in the cyclotron line
forming region is with the value inferred from the large scale dipole
field it if difficult to choose any one model over the others here. Such a choice can only be made based on the application of the models to a large sample of outbursts of different sources.

\subsection{Time dependence of magnetospheric radius}
\label{sec:timedep}

\begin{figure*}
\includegraphics[width=\hsize]{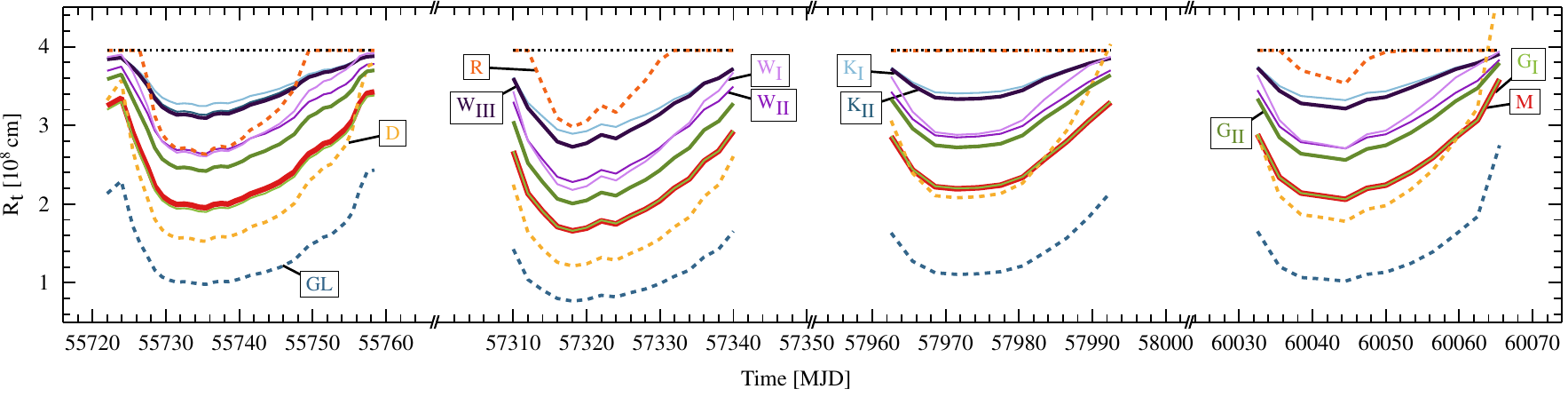}
\caption{Time dependence of the truncation radius for the
unconstrained best fits for each model. The labels indicate the 
model corresponding to the predicted truncation radius. Dashed
lines indicate models with $R_\tn{t} \propto R_\tn{A}$. The
dotted black line marks the radius of co-rotation.}
\label{fig:time-trunc}
\end{figure*}

From the determined fastness of the system we can also readily calculate the
expected location of the inner edge of the accretion disk. In Fig.~\ref{fig:time-trunc}
we show the truncation radius for all models over the outbursts, derived
from the unconstrained best fit solution with canonical parameters $M =
1.4\,M_\odot$ and $R = 10^6$\,cm. The figure also explains why the
resulting magnetic field distribution for the model of
\citet{rappaport2004a} discussed above has sharp edges: The fix-point at $\omega=1$
introduces sharp transitions in the truncation radius and are
therefore likely not well explored by the MCMC method. For the other
models, the absolute position of the truncation radius varies in
accordance to the hierarchy classes for the magnetic field estimates.
From the resulting distributions of the magnetic field
(Fig.~\ref{fig:magnetic}), it is evident that even the same models give
different results for the magnetic field for different outbursts. As
it is not expected that the large scale magnetic field changes
significantly over the observed time span, this is likely an artifact
of the model approach in general. One contribution certainly can be
attributed to the model for the orbit as it is to some extent in
conflict with the torque model. This is manifest in the slight variation
of the measured orbital parameters (see Appendix~\ref{apx:mcmc}).

A factor that needs to be considered is the quality of the data. The
measurement density of the spin period as well as of the light-curve
drive the result. This is most noticeable in the data for the 2023
outburst where nominally the magnetic field is most constrained.
However, this is only because the models overall do not fit the data
very well. The detailed interplay between the data quality and
applicability need to be studied further to resolve those problems.

\section{Other aspects of accretion and angular momentum change}
\label{sec:other-aspects}

All of the models considered here are based on simplifying assumptions
that are likely not realized in reality. First and foremost, all models
assume that accretion happens via a Keplerian disk with the angular
momentum aligned with the rotation axis and the magnetic dipole axis of
the accretor. For a more realistic picture, where at least one of those
symmetries are broken, additional effects of the disk-field interaction
might become relevant. Systems where accretion does not happen via a disk
might have very different behavior than predicted from the torque models
above which leads to further uncertainty in the measurements. Especially
relevant for evolutionary studies are other mechanisms that change the
angular momentum of the accretor. In this section we mention some of the
undertaken efforts that have been done to estimate effects due to additional
interactions or geometries, all of which are necessary to understand for
a more unified picture of the spin evolution of compact accretors.

\subsection{Inclination of magnetic axis}

The  highly symmetric case of the ``aligned rotator'' simplifies
calculations, but is also very 
unlikely to occur in reality. For an accreting neutron star it would not
even be consistent with pulsed emission that allows to measure the
spin period. Without changing the overall arguments for the accretion
process one can consider an inclined magnetic axis. In polar coordinates
$(r,\phi)$, the magnetic field in the disk plane for an inclined dipole
is given by \citep[e.g.,][]{jetzer1998a}
\begin{equation}
B^2 = \frac{\mu^2}{r^6}(1+3(\sin\chi\sin\phi)^2)
\end{equation}
with the inclination angle $\chi$. With this the truncation radius
is \citep{jetzer1998a,perna2006a}
\begin{equation}
R_\tn{t}(\phi)_\chi = R_\tn{t}(1+3(\sin\chi\sin\phi))^{2/7}\,.
\end{equation}
To first approximation the modification of the radius can be applied
to the Alfv\'en radius and the truncation radius is obtained for this
modified $R_\tn{A}$ now dependent on $\phi$. As a consequence one finds
that the truncation radius is at varying distance to the co-rotation
radius around the disk. From this it is expected that the plasma has a
different angular momentum depending on $\phi$. To get the total torque
on the central object one therefore has to integrate over $\phi$
taking this variation into account.

\subsection{Mass pileup at magnetosphere}
\label{sec:pileup}

Similar to the $\phi$ dependence of the magnetosphere, the general
torque will also depend on the structure of the disk at the truncation
radius in $z$ direction, that is, the height above the accretion disk.
The influence of this is explored by \citet{eksi2011a} and
\citet{cikintoglu2023a} based on its influence on the mass flow rate.
These authors argue that the magnetic field prevents matter from
accretion in the disk plane because there the plasma is forced to move in
the $z$ direction. Therefore the plasma piles up towards
larger $z$ until it reaches a point where it can penetrate the field.
This behavior has two consequences: first, it modifies the inferred
mass accretion rate at the magnetosphere from the luminosity, and
second, it introduces a natural asymmetry in the transition from
inhibited accretion to accretion and vice versa.

The key property that drives the evolution here is where the
magnetospheric boundary crosses the equilibrium surface. This surface
is defined by the equality of apparent acceleration due to the
centrifugal effect and the gravitational force along the magnetic
field lines. If at some distance $R_\tn{t}$ crosses inside the equilibrium
surface matter can penetrate and is accreted from above the crossing
point. If the surfaces do not cross, accretion is inhibited, leading to
the propeller state.

\subsection{Propeller state}
\label{sec:propeller}

To explain the synchronous orbits and periods of close X-ray binaries,
\citet{davidson1973a} introduced the concept of an equilibrium period
controlled by a centrifugal barrier possibly flinging material out of
the system. Later, and introducing the term propeller state,
\citet{illarionov1975a} used this accretion regime to explain the low
number of observed accreting neutron stars. The conceptual idea is
simple: If the accreted material couples to the magnetosphere at a
distance that is larger than the co-rotation radius, the matter is
accelerated onto trajectories that correspond to orbits with
increasing radius (where $R_\tn{t}$ is the closest point). Therefore,
for sufficiently extended magnetospheres matter will be pushed out of
the system and not accreted.

There is, however, a possible accretion regime where the centrifugal
effect does neither permit accretion onto the compact object, nor does
it provide enough momentum for the matter to leave the gravitational
potential. A simple ballistic argument gives a quantitative range for
the relevant radii for where this trapping state should occur
\citep{perna2006a}. In both cases angular momentum is removed from the
central object and converted either into momentum of the gas outflow
or convective motion.

The disk accretion models discussed above do not include the
transition from accretion over the trapped state to the propeller.
The quantification of the magnetosphere in those models via the mass
accretion rate introduces an artifact for this transition. We like to remind that
the magnetosphere is formally defined by the energy density balance
between the gas and the magnetic field. The energy density of the gas
can be expressed via its pressure, which in turn can be formally
expressed by the accretion rate. The pressure inward, however, is
counteracted by the pressure of the magnetic field. In the propeller
state the accretion rate is necessarily zero, as is the total pressure
at the magnetosphere. The pressure of the gas that determines the location
of the magnetosphere, however, is finite \citep{shakura2012a}.

\subsection{(Quasi-)spherical accretion}
\label{sec:quasisphere}

While the formation of an accretion disk is a very universal
phenomenon in astrophysical objects where the accreted flow has a
specific angular momentum, in binary systems where the donor star has
a strong a stellar wind with high velocity it is expected that instead a
shell of plasma forms around the accretor.
\citet{shakura2012a} explored the case for subsonic quasi-spherical
accretion, extending the discussion of \citet{davidson1973a} and
\citet{illarionov1975a}. With otherwise very similar arguments
compared to the disk case, \citeauthor{shakura2012a} find that the
magnetosphere scales differently with the mass accretion
rate compared to the disk models. In this model the wind and the angular
momentum of the plasma shell are formulated via the orbital parameters with
the wind velocity expressed as a function of the separation of the two bodies.
As a consequence the model intrinsically depends on the orbital separation
and orbital period, compared to the added modulation due to observer effects
(see Sect.~\ref{sec:example}).

The focus of the model of spherical accretion is the magnetospheric
boundary for a plasma shell. The dispersion velocity of the plasma at
the boundary is investigated to calculate an approximation of
$R_\tn{t}$ (labeled $R_\tn{A}$ by \citealt{shakura2012a}). The solution
is then written with a numerical factor, $f(u)$, that quantifies the
dispersion velocity via the free-fall velocity at $R_\tn{t}$. The
resulting relation between $R_\tn{t}$ and $R_\tn{A}$ then depends on
$(\mu/\dot{M})$. This additional factor prevents us from rewriting
this model solely as a function of the fastness and canonical
fastness.

It should be noted, however, that the expression of the model by
\citet{shakura2012a} is derived based on a linear approximation of
$f(u)$. This approximation improves for $\dot{M} \rightarrow 0$. On
the other hand, $f(u)$ must be $< 1$, which is not automatically
guaranteed in the linearized formulation of the
model. In a way, this is a similar problem to the disk models given
above for $\omega \rightarrow 1$.

\subsection{Super-Eddington accretion}
\label{sec:supereddi}

Another modification of the torque models happens for extremely large
mass accretion rates. These accretion regimes have gained a lot of attention
with the discovery of the first pulsating ultra-luminous x-ray source
\citep{bachetti2014a}. These sources are interesting for a number of reasons
and would benefit significantly if torquing can provide an estimate of the
magnetic field of the accretor. Due to the large amounts of matter that fall
towards the compact object the disk is no longer gas-pressure dominated but by
the radiation pressure. This causes a change in the disc structure, increasing
its height significantly \citep[e.g.,][]{chashkina2017a,mushtukov2019a}.
This change also affects the coupling between the matter and the magnetic field
such that the magnetospheric radius is altered \citep{chashkina2019a}.

From the theoretical descriptions of the change with increasing accretion rate
an effective model was constructed and applied to the galactic pulsating ultra-luminous
x-ray source Swift J0243.6+6124 \citep{atel:kennea2017a,atel:jenke2017a}. It was
found that the predicted effect for high accretion rates is clearly visible in
the data and that a model that takes this change into account is favored over
a model with $R_\tn{t} \sim R_\tn{A}$ \citep{karaferias2023a}. This change
can be interpreted as a change from a disk accreting state to one that corresponds to spherical accretion.

\subsection{Transition between accretion regimes}
\label{sec:accregion-regimes-connected}

In highly magnetized neutron star binaries it is suspected that during
an outburst some system undergo a transition from the propeller state
to the accretion state. From the asymptotic behavior of all of the
discussed torque models at $\omega = 1$ \citep[except][]{rappaport2004a},
it is clear that they do not work for this transition.
A general description of this transition has not yet been made, however, an
attempt to connect different accretion regimes by simple arguments was
given by \citet{ertan2020a}. The main argument here is that the
solution for the magnetospheric radius should also hold for $\omega
>1$. This solution branch can be obtained for all models where $R_\tn{t}$
is derived from the equilibrium condition\footnote{Note that
\citet{ertan2020a} derive their own estimate for $R_\tn{t}$ in a way that
does not allow for a description via $\omega$ and $\hat\omega$.}.
In the equilibrium condition only the absolute value of $\omega - 1$ is
meaningful, therefore replacing it with $|\omega-1|$ allows one to obtain
the second solution for $\omega>1$. The difference of the model of \citeauthor{ertan2020a} is
that there are two distinct solutions in the range for $\omega < 1$ and only one for $\omega
> 1$ while the situation is reversed for all models considered in our work.

\citet{ertan2020a} further argue that the second solution is only
valid for the case where matter can be effectively removed from the
system. They chose a ballistic solution similar to that obtained by
\citet{perna2006a} as the transition point ($R_\tn{t} >
2^{1/3}R_\tn{co}$). For smaller truncation radii, but still with
$\omega > 1$, they assume that truncation happens at $R_\tn{co}$
\citep{rappaport2004a}. To connect the two branches,
\citeauthor{ertan2020a} choose to intersect the truncation radius with
the linear solution of \citet{ghosh1979b} for a parametric scaling
factor. This approach, while not rigorous, allows one to investigate full
evolutionary tracks of accreting objects given initial parameters and
mass accretion and the influence of the various regimes.

\subsection{Additional loss of angular momentum}
\label{sec:additional-loss}

Besides the torque exchange through the linked disk there are other
contributions relevant for the spin evolution of a compact object.
For radio pulsars angular momentum is extracted by the pulsar
wind, caused by the rotating magnetic field, which steadily slows the
rotation \citep{pacini1967a, gold1968a, goldreich1969a}. For accreting
millisecond pulsars, \citet{parfrey2016a} derived a toy model, taking
the disk-field interaction as well as the pulsar wind into account.
The characteristic scale for this model is given by the
light-cylinder, $R_\tn{lc}$, which introduces an absolute value for
the characteristic radii. Due to this change of scale, the description
via the fastness contains an additional dependency on the rotation
period and can no longer be described via $\omega$ or $\hat\omega$.



\section{Conclusion}
\label{sec:conclusion}

In this paper, we showed that it is possible to express most models for the torque exerted onto a neutron star by its accretion flow in terms of one physical quantity, the canonical fastness, $\hat\omega$. Using this quantity ensures that the models are expressed with one set of physical quantities. Using $\hat\omega$, it is possible to directly compare many of the torque models in the literature, while comparing them using the model-dependent truncation radius hides the fact that the same truncation radius is realized for incompatible accretion scenarios ($\dot{M}$, $\mu$). We showed that the models generally follow similar trends and are much more similar than what is typically assumed when comparing their behavior using the model-dependent fastness, $\omega$. We then developed a general torquing formalism, showing that it is possible to describe many of the proposed simplified disk torque models with two parameters, $\cal{A}$ and $\cal{B}$, that can be related to the physical parameters of the accreting system. 

Despite the simplicity of the models, they are very successful in describing the observed data of the Be X-ray binary system 4U\,0115$+$63. The predicted system parameters, however, are far from consistent between outbursts, even when applying the same model. Considering that different and equally valid torquing parameterizations give results ranging over one order of magnitude, and given that no model yields parameters that are clearly statistically preferred argues against using  torque models to derive reliable physical properties of neutron stars at this stage. In particular, the derived magnetic field strength ranges over one order of magnitude between models, even without accounting for the systematic uncertainty in the interaction parameters, which would increase the uncertainty of the model parameters even more. As an example, for the Galactic ultra-luminous x-ray pulsar Swift~J$0243.6+6124$ \citep{kong2022a} argue that the magnetic field inferred from its cyclotron line, $B \approx 1.6\times 10^{13}$\,G is in contradiction with the results from magnetic field estimates based on torque theory \citep[e.g.,][]{jaisawal2019a}. However, given the extreme $\dot{M}$ of this source, in combination with a large $B$-field strength, estimated from cyclotron line measurements, the system should be considered ``fast''. This is exactly the range that is affected most strongly by the linear approximation of \citet{ghosh1979b}, making torque theory unreliable. In a similar fashion, in their simplified form torquing models are generally not consistent with other methods estimating the extent of the magnetosphere \citep{bozzo2009a}.

Despite their shortcomings, however, these simplified torquing models are still useful to understand general trends of the spin period for disk accretors, linking $\dot{M}$ to the spin period. With the latter being much better constrained by observations and combining this with physical spectral models it might be possible to invert the problem. By this we mean that instead of using the mass accretion rate (inferred from the light-curve) as input, one could use the observed pulse period as a proxy for the mass accretion rate parameter used by some physical accretion models \citep[such as][]{becker2007a,sokolovalapa2021a}. The drawback of such an approach would be that it requires access to a number of high quality broad-band X-ray spectra covering a large part of an outburst, which might not be easy to obtain.

A systematic study of proposed torque models and their application to the rich dataset observed by individual observer campaigns and monitoring programs such as from BATSE \citep{bildsten1997a} and Fermi/GBM \citep{malacaria2020a} will allow us to understand and correct for the systematics in the analytical torque models. The comparison with data and also with numerical simulations might provide a way to connect the accretion process to the low luminosity or propeller state. As mentioned before, at low $\dot{M}$ luminosity is no longer a good proxy for $\dot{M}$ because it appears only formally in the equations, representing the pressure at $R_\tn{t}$. In the transition to a non-accreting regime $\dot{M}$ approaches zero while the pressure stays finite. Solving this problem in the torque models can provide additional constraints to understand systematic deviations.

Given the simplicity of most torquing models it is not particularly surprising that the results are inconsistent or biased. Still, identifying a model that at least approximately yields somewhat trustworthy magnetic field estimates would be very useful to compare different accretors and their accretion stages. In the particular case of 4U0115+63 discussed here, at least order of magnitude estimates seem to be consistent with cyclotron line measurements. Understanding of whether this holds in general could be achieved by a broader study of CSRF sources, describing their pulse period evolution with different torquing models and then identifying  the model which is most consistent with the magnetic field determined from cyclotron lines. There are several caveats, though. First, it is not clear whether the CSRF forming region is a good representation of the large scale magnetic field. And second, different sources might be best described by different torquing models.

To conclude, it appears that applications of torque theory in order to estimate the magnetic field of accreting neutron stars are more strongly affected by systematic uncertainties than generally acknowledged in the literature. Torquing theory can explain general trends in the pulse period behavior, but in its current state it is unreliable and does not yield precise numbers.

\begin{acknowledgements}
This research was supported by the
International Space Science Institute (ISSI) in Bern,
through ISSI International Team project \#495 (Feeding the
spinning top). 
GV acknowledges support from the Hellenic Foundation for Research and Innovation (H.F.R.I.) through the project ASTRAPE (Project ID 7802).
CM acknowledges funding from the Italian Ministry of University and
Research (MUR), PRIN 2020 (prot. 2020BRP57Z)
\textit{``Gravitational and Electro magnetic-wave Sources in the Universe
with current and next generation
detectors (GEMS)''} and the INAF Research Grant \textit{Uncovering the
optical beat of the fastest magnetised neutron stars (FANS)}.
JW and ESL acknowledge support from Deutsche Forschungsgemeinschaft grant WI 1860/11-2.
\end{acknowledgements}


\bibliographystyle{aa}
\bibliography{ref.bib}

\begin{thebibliography}{96}
\expandafter\ifx\csname natexlab\endcsname\relax\def\natexlab#1{#1}\fi

\bibitem[{Alfv{\'{e}}n \& Lindblad(1947)}]{alfven1947a}
Alfv{\'{e}}n, H. \& Lindblad, B. 1947, \mnras, 107, 211

\bibitem[{Bachetti {et~al.}(2014)Bachetti, Harrison, Walton, Grefenstette,
  Chakrabarty, F{\"u}rst, Barret, Beloborodov, Boggs, Christensen, Craig,
  Fabian, Hailey, Hornschemeier, Kaspi, Kulkarni, Maccarone, Miller, Rana,
  Stern, Tendulkar, Tomsick, Webb, \& Zhang}]{bachetti2014a}
Bachetti, M., Harrison, F.~A., Walton, D.~J., {et~al.} 2014, \nat, 514, 202

\bibitem[{Basko \& Sunyaev(1976)}]{basko1976b}
Basko, M.~M. \& Sunyaev, R.~A. 1976, \sovast, 20, 537

\bibitem[{{Becker} {et~al.}(2012){Becker}, {Klochkov}, {Sch{\"{o}}nherr},
  {Nishimura}, {Ferrigno}, {Caballero}, {Kretschmar}, {Wolff}, {Wilms}, \&
  {Staubert}}]{becker2012a}
{Becker}, P.~A., {Klochkov}, D., {Sch{\"{o}}nherr}, G., {et~al.} 2012, \aap,
  544, A123

\bibitem[{Becker \& Wolff(2007)}]{becker2007a}
Becker, P.~A. \& Wolff, M.~T. 2007, \apj, 654, 435

\bibitem[{Bildsten {et~al.}(1997)Bildsten, Chakrabarty, Chiu, Finger, Koh,
  Nelson, Prince, Rubin, Scott, Stollberg, Vaughan, Wilson, \&
  Wilson}]{bildsten1997a}
Bildsten, L., Chakrabarty, D., Chiu, J., {et~al.} 1997, \apjs, 113, 367

\bibitem[{{Bissinger} {et~al.}(2020){Bissinger}, {Kreykenbohm}, {Ferrigno},
  {Pottschmidt}, {Marcu-Cheatham}, {F\"urst}, {Rothschild}, {Kretschmar},
  {Klochkov}, {Hemphill}, {Hertel}, {M\"uller}, {Sokolova-Lapa}, {Oruru},
  {Grinberg}, {Mart\'{\i}nez-N\'u\~nez}, {Torrej\'on}, {Becker}, {Wolff},
  {Ballhausen}, {Schwarm}, \& {Wilms}}]{kuehnel2020a}
{Bissinger}, n\'e~K\"uhnel, M., {Kreykenbohm}, I., {Ferrigno}, C., {et~al.}
  2020, \aap, 634, A99

\bibitem[{Blondin(2013)}]{blondin2013a}
Blondin, J.~M. 2013, \apj, 767, 135

\bibitem[{Bondi \& Hoyle(1944)}]{bondi1944a}
Bondi, H. \& Hoyle, F. 1944, \mnras, 104, 273

\bibitem[{Bozzo {et~al.}(2008)Bozzo, Falanga, \& Stella}]{bozzo2008a}
Bozzo, E., Falanga, M., \& Stella, L. 2008, \apj, 683, 1031

\bibitem[{{Bozzo} {et~al.}(2009){Bozzo}, {Stella}, {Vietri}, \&
  {Ghosh}}]{bozzo2009a}
{Bozzo}, E., {Stella}, L., {Vietri}, M., \& {Ghosh}, P. 2009, \aap, 493, 809

\bibitem[{Campana(2001)}]{campana2001a}
Campana, S. 2001, AIP Conf. Proc., 599, 63

\bibitem[{{Campbell}(2018)}]{book:campbell2018a}
{Campbell}, C.~G. 2018, {Magnetohydrodynamics in Binary Stars} (Springer)

\bibitem[{{{\c{C}}}{\i}k{\i}nto{\u{g}}lu \& Ek{\c{s}}i(2023)}]{cikintoglu2023a}
{{\c{C}}}{\i}k{\i}nto{\u{g}}lu, S. \& Ek{\c{s}}i, K.~Y. 2023, \mnras, 524, 1727

\bibitem[{Chashkina {et~al.}(2017)Chashkina, Abolmasov, \&
  Poutanen}]{chashkina2017a}
Chashkina, A., Abolmasov, P., \& Poutanen, J. 2017, \mnras, 470, 2799

\bibitem[{Chashkina {et~al.}(2019)Chashkina, Lipunova, Abolmasov, \&
  Poutanen}]{chashkina2019a}
Chashkina, A., Lipunova, G., Abolmasov, P., \& Poutanen, J. 2019, \aap, 626,
  A18

\bibitem[{Cropper(1990)}]{cropper1990a}
Cropper, M. 1990, \ssr, 54, 195

\bibitem[{{Dai} \& {Li}(2006)}]{dai2006a}
{Dai}, H.-L. \& {Li}, X.-D. 2006, \aap, 451, 581

\bibitem[{{Davidson} \& {Ostriker}(1973)}]{davidson1973a}
{Davidson}, K. \& {Ostriker}, J.~P. 1973, \apj, 179, 585

\bibitem[{Edgar(2004)}]{edgar2004a}
Edgar, R. 2004, \nar, 48, 843

\bibitem[{Ekşi \& Kutlu(2011)}]{eksi2011a}
Ekşi, K.~Y. \& Kutlu, E. 2011, AIP Conf. Proc., 1379, 156

\bibitem[{El~Mellah {et~al.}(2019{\natexlab{a}})El~Mellah, Sander, Sundqvist,
  \& Keppens}]{elmellah2019a}
El~Mellah, I., Sander, A. A.~C., Sundqvist, J.~O., \& Keppens, R.
  2019{\natexlab{a}}, \aap, 622, A189

\bibitem[{El~Mellah {et~al.}(2019{\natexlab{b}})El~Mellah, Sundqvist, \&
  Keppens}]{elmellah2019b}
El~Mellah, I., Sundqvist, J.~O., \& Keppens, R. 2019{\natexlab{b}}, \aap, 622,
  L3

\bibitem[{{Elsner} \& {Lamb}(1977)}]{elsner1977a}
{Elsner}, R.~F. \& {Lamb}, F.~K. 1977, \apj, 215, 897

\bibitem[{Ertan(2020)}]{ertan2020a}
Ertan, {\"{U}}. 2020, \mnras, 500, 2928

\bibitem[{{Farinelli, Ruben} {et~al.}(2016){Farinelli, Ruben}, {Ferrigno,
  Carlo}, {Bozzo, Enrico}, \& {Becker, Peter A.}}]{farinelli2016a}
{Farinelli, Ruben}, {Ferrigno, Carlo}, {Bozzo, Enrico}, \& {Becker, Peter A.}
  2016, \aap, 591, A29

\bibitem[{Foreman-Mackey {et~al.}(2013)Foreman-Mackey, Hogg, Lang, \&
  Goodman}]{foremanmackey2013a}
Foreman-Mackey, D., Hogg, D.~W., Lang, D., \& Goodman, J. 2013, PASP, 125, 306

\bibitem[{Frank {et~al.}(2002)Frank, King, \& Raine}]{book:frank2002a}
Frank, J., King, A., \& Raine, D. 2002, {A}ccretion {P}ower in {A}strophysics,
  3rd edn. (Cambridge University Press)

\bibitem[{Gao \& Li(2021)}]{gao2021a}
Gao, S.-J. \& Li, X.-D. 2021, Res. Astron. Astrophys., 21, 196

\bibitem[{Gehrels {et~al.}(2005)Gehrels, Chincarini, Giommi, Mason, Nousek,
  Wells, White, Barthelmy, Burrows, Cominsky, Hurley, Marshall,
  M{\'{e}}sz{\'{a}}ros, Roming, Angelini, Barbier, Belloni, Boyd, Campana,
  Caraveo, Chester, Citterio, Cline, Cropper, Cummings, Dean, Feigelson,
  Fenimore, Frail, Fruchter, Garmire, Gendreau, Ghisellini, Greiner, Hill,
  Hunsberger, Krimm, Kulkarni, Kumar, Lebrun, Lloyd-Ronning, Markwardt,
  Mattson, Mushotzky, Norris, Paczynski, Palmer, Park, Parsons, Paul, Rees,
  Reynolds, Rhoads, Sasseen, Schaefer, Short, Smale, Smith, Stella, Still,
  Tagliaferri, Takahashi, Tashiro, Townsley, Tueller, Turner, Vietri, Voges,
  Ward, Willingale, Zerbi, \& Zhang}]{gehrels2005a}
Gehrels, N., Chincarini, G., Giommi, P., {et~al.} 2005, \apj, 621, 558

\bibitem[{Gehrels {et~al.}(2004)Gehrels, Chincarini, Giommi, Mason, Nousek,
  Wells, White, Barthelmy, Burrows, Cominsky, Hurley, Marshall,
  M{\'{e}}sz{\'{a}}ros, Roming, Angelini, Barbier, Belloni, Campana, Caraveo,
  Chester, Citterio, Cline, Cropper, Cummings, Dean, Feigelson, Fenimore,
  Frail, Fruchter, Garmire, Gendreau, Ghisellini, Greiner, Hill, Hunsberger,
  Krimm, Kulkarni, Kumar, Lebrun, Lloyd-Ronning, Markwardt, Mattson, Mushotzky,
  Norris, Osborne, Paczynski, Palmer, Park, Parsons, Paul, Rees, Reynolds,
  Rhoads, Sasseen, Schaefer, Short, Smale, Smith, Stella, Tagliaferri,
  Takahashi, Tashiro, Townsley, Tueller, Turner, Vietri, Voges, Ward,
  Willingale, Zerbi, \& Zhang}]{gehrels2004a}
Gehrels, N., Chincarini, G., Giommi, P., {et~al.} 2004, \apj, 611, 1005

\bibitem[{{Ghosh} \& {Lamb}(1979)}]{ghosh1979b}
{Ghosh}, P. \& {Lamb}, F.~K. 1979, \apj, 232, 259

\bibitem[{Ghosh {et~al.}(1977)Ghosh, Lamb, \& Pethick}]{ghosh1977a}
Ghosh, P., Lamb, F.~K., \& Pethick, C.~J. 1977, \apj, 217, 578

\bibitem[{Giacconi {et~al.}(1972)Giacconi, Murray, Gursky, Kellogg, Schreier,
  \& Tananbaum}]{giacconi1972a}
Giacconi, R., Murray, S., Gursky, H., {et~al.} 1972, \apj, 178, 281

\bibitem[{Gold(1968)}]{gold1968a}
Gold, T. 1968, Nature, 218, 731

\bibitem[{Goldreich \& Julian(1969)}]{goldreich1969a}
Goldreich, P. \& Julian, W.~H. 1969, \apj, 157, 869

\bibitem[{Goodson {et~al.}(1999)Goodson, B{\"{o}}hm, \& Winglee}]{goodson1999a}
Goodson, A.~P., B{\"{o}}hm, K.-H., \& Winglee, R.~M. 1999, \apj, 524, 142

\bibitem[{Goodson \& Winglee(1999)}]{goodson1999b}
Goodson, A.~P. \& Winglee, R.~M. 1999, \apj, 524, 159

\bibitem[{Hartmann(1999)}]{hartmann1999a}
Hartmann, L. 1999, \nar, 43, 1

\bibitem[{Hayasaki \& Okazaki(2004)}]{hayasaki2004a}
Hayasaki, K. \& Okazaki, A.~T. 2004, MNRAS, 350, 971

\bibitem[{{Illarionov} \& {Sunyaev}(1975)}]{illarionov1975a}
{Illarionov}, A.~F. \& {Sunyaev}, R.~A. 1975, \aap, 39, 185

\bibitem[{Ireland {et~al.}(2022)Ireland, Matt, \& Zanni}]{ireland2022a}
Ireland, L.~G., Matt, S.~P., \& Zanni, C. 2022, \apj, 929, 65

\bibitem[{Ireland {et~al.}(2020)Ireland, Zanni, Matt, \&
  Pantolmos}]{ireland2020a}
Ireland, L.~G., Zanni, C., Matt, S.~P., \& Pantolmos, G. 2020, \apj, 906, 4

\bibitem[{Jaisawal {et~al.}(2019)Jaisawal, Wilson-Hodge, Fabian, Naik,
  Chakrabarty, Kretschmar, Ballantyne, Ludlam, Chenevez, Altamirano,
  Arzoumanian, F{\"{u}}rst, Gendreau, Guillot, Malacaria, Miller, Stevens, \&
  Wolff}]{jaisawal2019a}
Jaisawal, G.~K., Wilson-Hodge, C.~A., Fabian, A.~C., {et~al.} 2019, \apj, 885,
  18

\bibitem[{{Jenke} \& {Wilson-Hodge}(2017)}]{atel:jenke2017a}
{Jenke}, P. \& {Wilson-Hodge}, C.~A. 2017, The Astronomer's Telegram, 10812, 1

\bibitem[{Jetzer {et~al.}(1998)Jetzer, Str{\"{a}}ssle, \&
  Straumann}]{jetzer1998a}
Jetzer, P., Str{\"{a}}ssle, M., \& Straumann, N. 1998, \na, 3, 619

\bibitem[{Karaferias {et~al.}(2023)Karaferias, Vasilopoulos, Petropoulou,
  Jenke, Wilson-Hodge, \& Malacaria}]{karaferias2023a}
Karaferias, A.~S., Vasilopoulos, G., Petropoulou, M., {et~al.} 2023, \mnras,
  520, 281

\bibitem[{{Kennea} {et~al.}(2017){Kennea}, {Lien}, {Krimm}, {Cenko}, \&
  {Siegel}}]{atel:kennea2017a}
{Kennea}, J.~A., {Lien}, A.~Y., {Krimm}, H.~A., {Cenko}, S.~B., \& {Siegel},
  M.~H. 2017, ATel, 10809, 1

\bibitem[{{Klement} {et~al.}(2017){Klement}, {Carciofi}, {Rivinius},
  {Matthews}, {Vieira}, {Ignace}, {Bjorkman}, {Mota}, {Faes}, {Bratcher},
  {Cur{\'e}}, \& {{\v{S}}tefl}}]{klement2017a}
{Klement}, R., {Carciofi}, A.~C., {Rivinius}, T., {et~al.} 2017, \aap, 601, A74

\bibitem[{Klu{\'{z}}niak \& Rappaport(2007)}]{kluzniak2007a}
Klu{\'{z}}niak, W. \& Rappaport, S. 2007, \apj, 671, 1990

\bibitem[{Kong {et~al.}(2022)Kong, Zhang, Zhang, Ji, Doroshenko, Santangelo,
  Chen, Lu, Ge, Wang, Tao, Qu, Li, Liu, Liao, Chang, Peng, \& Shui}]{kong2022a}
Kong, L.-D., Zhang, S., Zhang, S.-N., {et~al.} 2022, The Astrophysical Journal
  Letters, 933, L3

\bibitem[{{Lamb} {et~al.}(1973){Lamb}, {Pethick}, \& {Pines}}]{lamb1973a}
{Lamb}, F.~K., {Pethick}, C.~J., \& {Pines}, D. 1973, \apj, 184, 271

\bibitem[{Larson(2003)}]{larson2003a}
Larson, R.~B. 2003, Rep. Prog. Phys., 66, 1651

\bibitem[{Lipunov(1987)}]{lipunov1987a}
Lipunov, V.~M. 1987, \apss, 132, 1

\bibitem[{Lipunov(1992)}]{book:lipunov1992a}
Lipunov, V.~M. 1992, {Astrophysics of Neutron Stars} (Springer Berlin,
  Heidelberg)

\bibitem[{Malacaria {et~al.}(2020)Malacaria, Jenke, Roberts, Wilson-Hodge,
  Cleveland, Mailyan, \& on~behalf of~the GBM Accreting Pulsars
  Program~Team}]{malacaria2020a}
Malacaria, C., Jenke, P., Roberts, O.~J., {et~al.} 2020, \apj, 896, 90

\bibitem[{Matt \& Pudritz(2005)}]{matt2005a}
Matt, S. \& Pudritz, R.~E. 2005, \mnras, 356, 167

\bibitem[{McKee \& Ostriker(2007)}]{mckee2007a}
McKee, C.~F. \& Ostriker, E.~C. 2007, \araa, 45, 565

\bibitem[{Meegan {et~al.}(2009)Meegan, Lichti, Bhat, Bissaldi, Briggs,
  Connaughton, Diehl, Fishman, Greiner, Hoover, van~der Horst, von Kienlin,
  Kippen, Kouveliotou, McBreen, Paciesas, Preece, Steinle, Wallace, Wilson, \&
  Wilson-Hodge}]{meegan2009a}
Meegan, C., Lichti, G., Bhat, P.~N., {et~al.} 2009, \apj, 702, 791

\bibitem[{Mukai(2017)}]{mukai2017a}
Mukai, K. 2017, PASP, 129, 062001

\bibitem[{Mushtukov {et~al.}(2019)Mushtukov, Ingram, Middleton, Nagirner, \&
  van{~}der{~}Klis}]{mushtukov2019a}
Mushtukov, A.~A., Ingram, A., Middleton, M., Nagirner, D.~I., \&
  van{~}der{~}Klis, M. 2019, \mnras, 484, 687

\bibitem[{Mushtukov {et~al.}(2015)Mushtukov, Suleimanov, Tsygankov, \&
  Poutanen}]{mushtukov2015a}
Mushtukov, A.~A., Suleimanov, V.~F., Tsygankov, S.~S., \& Poutanen, J. 2015,
  \mnras, 447, 1847

\bibitem[{Nathanail \& Contopoulos(2014)}]{nathanail2014a}
Nathanail, A. \& Contopoulos, I. 2014, \apj, 788, 186

\bibitem[{Okazaki {et~al.}(2002)Okazaki, Bate, Ogilvie, \&
  Pringle}]{okazaki2002a}
Okazaki, A.~T., Bate, M.~R., Ogilvie, G.~I., \& Pringle, J.~E. 2002, \mnras,
  337, 967

\bibitem[{Pacini(1967)}]{pacini1967a}
Pacini, F. 1967, Nature, 216, 567

\bibitem[{{Panoglou} {et~al.}(2016){Panoglou}, {Carciofi}, {Vieira}, {Cyr},
  {Jones}, {Okazaki}, \& {Rivinius}}]{panoglou2016a}
{Panoglou}, D., {Carciofi}, A.~C., {Vieira}, R.~G., {et~al.} 2016, \mnras, 461,
  2616

\bibitem[{Parfrey {et~al.}(2016)Parfrey, Spitkovsky, \&
  Beloborodov}]{parfrey2016a}
Parfrey, K., Spitkovsky, A., \& Beloborodov, A.~M. 2016, \apj, 822, 33

\bibitem[{{Perna} {et~al.}(2006){Perna}, {Bozzo}, \& {Stella}}]{perna2006a}
{Perna}, R., {Bozzo}, E., \& {Stella}, L. 2006, \apj, 639, 363

\bibitem[{Popham \& Narayan(1991)}]{popham1991a}
Popham, R. \& Narayan, R. 1991, \apj, 370, 604

\bibitem[{Postnov {et~al.}(2015)Postnov, Gornostaev, Klochkov, Laplace, Lukin,
  \& Shakura}]{postnov2015a}
Postnov, K.~A., Gornostaev, M.~I., Klochkov, D., {et~al.} 2015, \mnras, 452,
  1601

\bibitem[{Pringle \& Rees(1972)}]{pringle1972a}
Pringle, J.~E. \& Rees, M.~J. 1972, \aap, 21, 1

\bibitem[{Rappaport {et~al.}(2004)Rappaport, Fregeau, \&
  Spruit}]{rappaport2004a}
Rappaport, S.~A., Fregeau, J.~M., \& Spruit, H. 2004, \apj, 606, 436

\bibitem[{Reig(2011)}]{reig2011a}
Reig, P. 2011, \apss, 332, 1

\bibitem[{Rivinius {et~al.}(2013)Rivinius, Carciofi, \&
  Martayan}]{rivinius2013a}
Rivinius, T., Carciofi, A.~C., \& Martayan, C. 2013, \aapr, 21, 69

\bibitem[{Romanova {et~al.}(2018)Romanova, Blinova, Ustyugova, Koldoba, \&
  Lovelace}]{romanova2018a}
Romanova, M., Blinova, A., Ustyugova, G., Koldoba, A., \& Lovelace, R. 2018,
  \na, 62, 94

\bibitem[{Romanova {et~al.}(2008)Romanova, Kulkarni, \&
  Lovelace}]{romanova2008a}
Romanova, M.~M., Kulkarni, A.~K., \& Lovelace, R. V.~E. 2008, \apj, 673, L171

\bibitem[{Santangelo {et~al.}(1999)Santangelo, Segreto, Giarrusso, Dal~Fiume,
  Orlandini, Parmar, Oosterbroek, Bulik, Mihara, Campana, Israel, \&
  Stella}]{santangelo1999a}
Santangelo, A., Segreto, A., Giarrusso, S., {et~al.} 1999, \apj, 523, L85

\bibitem[{Shakura {et~al.}(2012)Shakura, Postnov, Kochetkova, \&
  Hjalmarsdotter}]{shakura2012a}
Shakura, N., Postnov, K., Kochetkova, A., \& Hjalmarsdotter, L. 2012, \mnras,
  420, 216

\bibitem[{{Shakura} \& {Sunyaev}(1973)}]{shakura1973a}
{Shakura}, N.~I. \& {Sunyaev}, R.~A. 1973, \aap, 24, 337

\bibitem[{Shklovsky(1967)}]{shklovsky1967a}
Shklovsky, I.~S. 1967, \apjl, 148, L1

\bibitem[{Shvartsman(1970)}]{shvartsman1970a}
Shvartsman, V.~F. 1970, Radiophys. Quantum Electron., 13, 1428

\bibitem[{Sokolova-Lapa(2023)}]{phd:sokolovalapa2023a}
Sokolova-Lapa, E. 2023, PhD thesis, Friedrich-Alexander Universität
  Erlangen-Nürnberg

\bibitem[{{Sokolova-Lapa} {et~al.}(2021){Sokolova-Lapa}, {Gornostaev}, {Wilms},
  {Ballhausen}, {Falkner}, {Postnov}, {Thalhammer}, {F{\"{u}}rst},
  {Garc{\'{i}}a}, {Shakura}, {Becker}, {Wolff}, {Pottschmidt}, {H{\"{a}}rer},
  \& {Malacaria}}]{sokolovalapa2021a}
{Sokolova-Lapa}, E., {Gornostaev}, M., {Wilms}, J., {et~al.} 2021, \aap, 651,
  A12

\bibitem[{{Staubert} {et~al.}(2019){Staubert}, {Tr\"umper}, {Kendziorra, E.},
  {Klochkov, D.}, {Postnov, K.}, {Kretschmar, P.}, {Pottschmidt, K.}, {Haberl,
  F.}, {Rothschild, R. E.}, {Santangelo, A.}, {Wilms, J.}, {Kreykenbohm, I.},
  \& {F\"urst, F.}}]{staubert2019a}
{Staubert}, R., {Tr\"umper}, J., {Kendziorra, E.}, {et~al.} 2019, \aap, 622,
  A61

\bibitem[{Syunyaev \& Shakura(1986)}]{syunyaev1986a}
Syunyaev, R.~A. \& Shakura, N.~I. 1986, Sov. Astron. Lett., 12, 117

\bibitem[{Toropina {et~al.}(2003)Toropina, Romanova, Toropin, \&
  Lovelace}]{toropina2003a}
Toropina, O.~D., Romanova, M.~M., Toropin, Y.~M., \& Lovelace, R. V.~E. 2003,
  \apj, 593, 472

\bibitem[{Uzdensky(2004)}]{uzdensky2004a}
Uzdensky, D.~A. 2004, \apj, 603, 652

\bibitem[{Uzdensky(2005)}]{uzdensky2005a}
Uzdensky, D.~A. 2005, \apj, 620, 889

\bibitem[{{Wang}(1987)}]{wang1987a}
{Wang}, Y.~M. 1987, \aap, 183, 257

\bibitem[{{Wang}(1995)}]{wang1995a}
{Wang}, Y.~M. 1995, \apjl, 449, L153

\bibitem[{West {et~al.}(2017)West, Wolfram, \& Becker}]{west2017a}
West, B.~F., Wolfram, K.~D., \& Becker, P.~A. 2017, \apj, 835, 130

\bibitem[{Wheaton {et~al.}(1979)Wheaton, Doty, Primini, Cooke, Dobson, Goldman,
  Hecht, Hoffman, Howe, Scheepmaker, Tsiang, Lewin, Matteson, Gruber, Baity,
  Rothschild, Knight, Nolan, \& Peterson}]{wheaton1979a}
Wheaton, W.~A., Doty, J.~P., Primini, F.~A., {et~al.} 1979, Nature, 282, 240

\bibitem[{Wheeler(1993)}]{book:wheeler1993a}
Wheeler, J.~C. 1993, Accretion Disks in Compact Stellar Systems (WORLD
  SCIENTIFIC)

\bibitem[{Yi(1995)}]{yi1995a}
Yi, I. 1995, \apj, 442, 768

\bibitem[{Zanni \& Ferreira(2013)}]{zanni2013a}
Zanni, C. \& Ferreira, J. 2013, \aap, 550, A99

\bibitem[{Zel'dovich \& Novikov(1966)}]{zeldovich1966a}
Zel'dovich, Y.~B. \& Novikov, I.~D. 1966, Sov. Phys. Uspekhi, 8, 522

\end{thebibliography}

\FloatBarrier
\begin{appendix}

\section{Monte Carlo Markov Chain results}
\label{apx:mcmc}

To explore the parameter space of the application of all considered
models to the data of the Be X-ray binary 4U\,0115$+$63, we used the
algorithm described by \citet{foremanmackey2013a}. The model includes
the torque parameters $\mathcal{A}$ and $\mathcal{B}$, together with
orbital parameters projected semi-major axis ($a\,\sin i$), orbital
period ($P_\tn{orb}$), argument of periapsis ($\omega$, not to be
confused with the fastness here), time of periastron passage
($\tau$), and eccentricity ($e$).

The torque parameters are expressed in a convenient form as
\begin{equation}
\begin{split}
\frac{\mathcal{A}}{\mathcal{A}_0} &=  \left(\frac{\mu}{10^{30}\,\tn{G}\,\tn{cm}^3} \right) \left(\frac{M}{M_\odot} \right)^{-1} \left(\frac{R}{10^6\,\tn{cm}} \right)^{-1}
\\
\frac{\mathcal{B}}{\mathcal{B}_0} &=  h^{-3/7}\left(\frac{\mu}{10^{30}\,\tn{G\,cm}^3}\right)^{6/7} \left(\frac{R}{10^6\,\tn{cm}} \right)^{-3/7} \left(\frac{M}{M_\odot} \right)^{-2/7} \left(\frac{D}{1\,\tn{kpc}} \right)^{-6/7} \\
\mathcal{A}_0 &\approx 10^{24}\,\tn{G\,cm}^2\,M_\odot^{-1} \\
\mathcal{B}_0 &\approx 3.59 \times 10^{37}\,\tn{kpc}^2\,\tn{g}^2\,\tn{cm}^{-1}\,.
\end{split}
\end{equation}

\begin{figure*}
\includegraphics[width=\hsize]{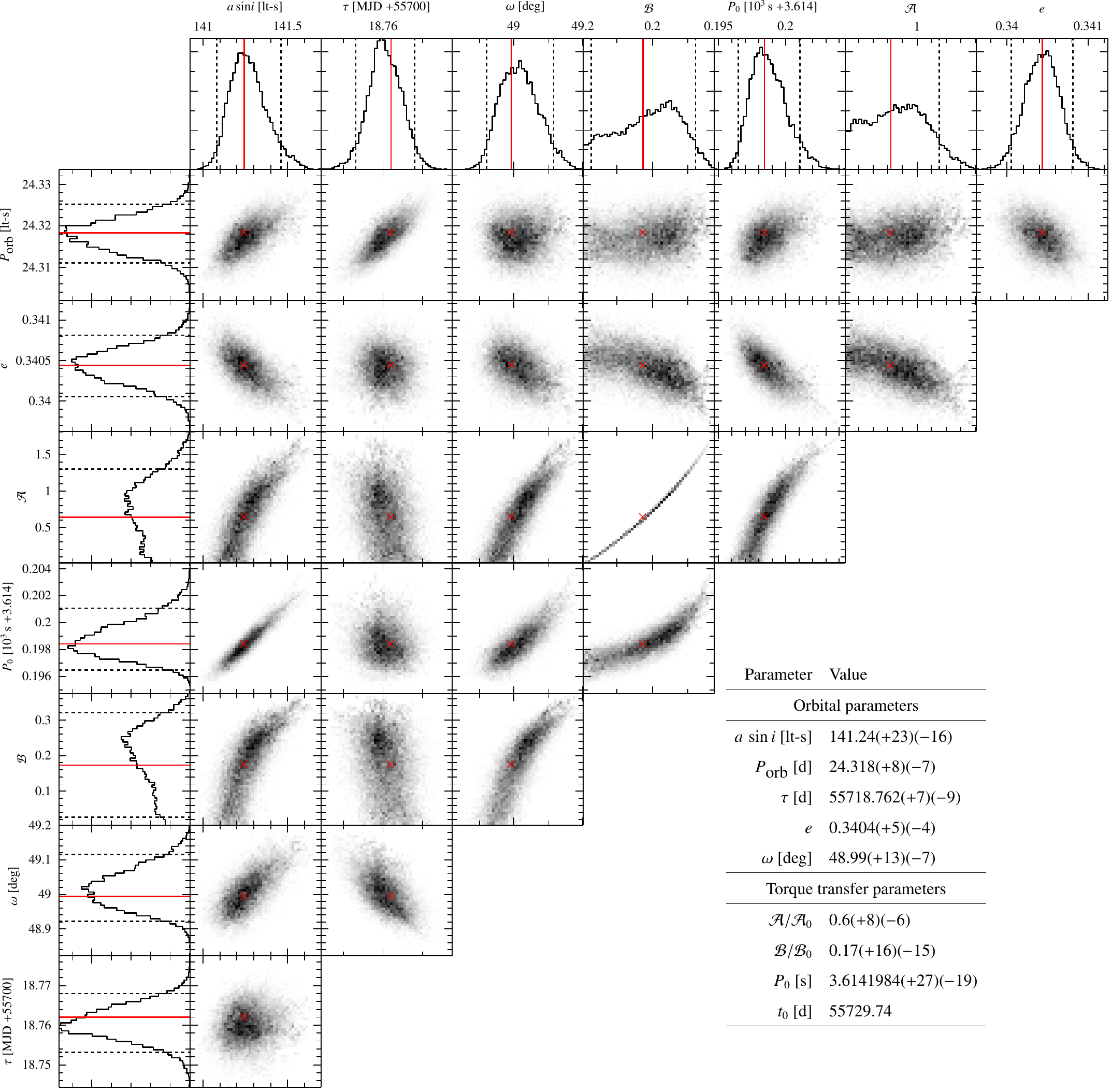}
\caption{Resulting parameter distribution of W$_\tn{III}$ for the 2011
outburst of 4U\,0115. Orbital distance is given in light-seconds,
orbital velocity is given in units of speed-of-light, standard
gravitational parameters is scaled for light-days and days, and
the reference time is given in days. For $\mathcal{A}$ and
$\mathcal{B}$ the units are omitted, values are given for a unit
system where $\mu \sim 10^{30}$\,G\,cm$^3$, $M \sim 1\,M_\odot$,
$R \sim 10^6$\,cm, $P \sim 1$\,s, and $D \sim 1$\,kpc. The dashed
lines in the marginal distributions indicate the 90\% confidence
interval. The red line indicates the best fit value in the
ensemble. The red crosses mark the corresponding best fit value
in the 2D projections. The table gives the corresponding values.
The number brackets indicate the 90\% uncertainty (symmetric for
one bracket).}
\label{fig:mcmc-1}
\end{figure*}

\begin{figure*}
\includegraphics[width=\hsize]{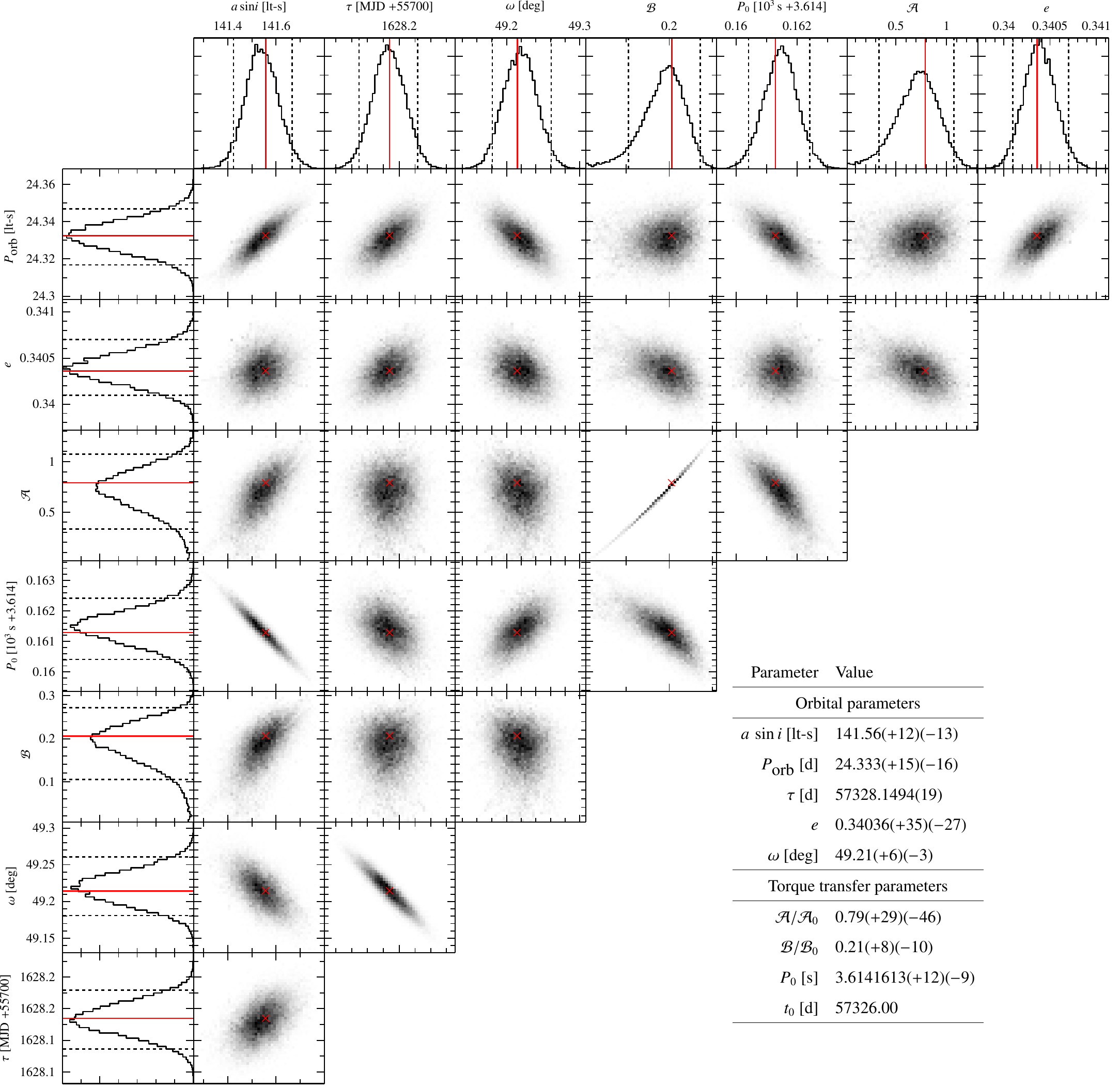}
\caption{Resulting parameter distribution of W$_\tn{III}$ for the 2015
outburst of 4U\,0115. See Fig.~\ref{fig:mcmc-1}.}
\label{fig:mcmc-2}
\end{figure*}

\begin{figure*}
\includegraphics[width=\hsize]{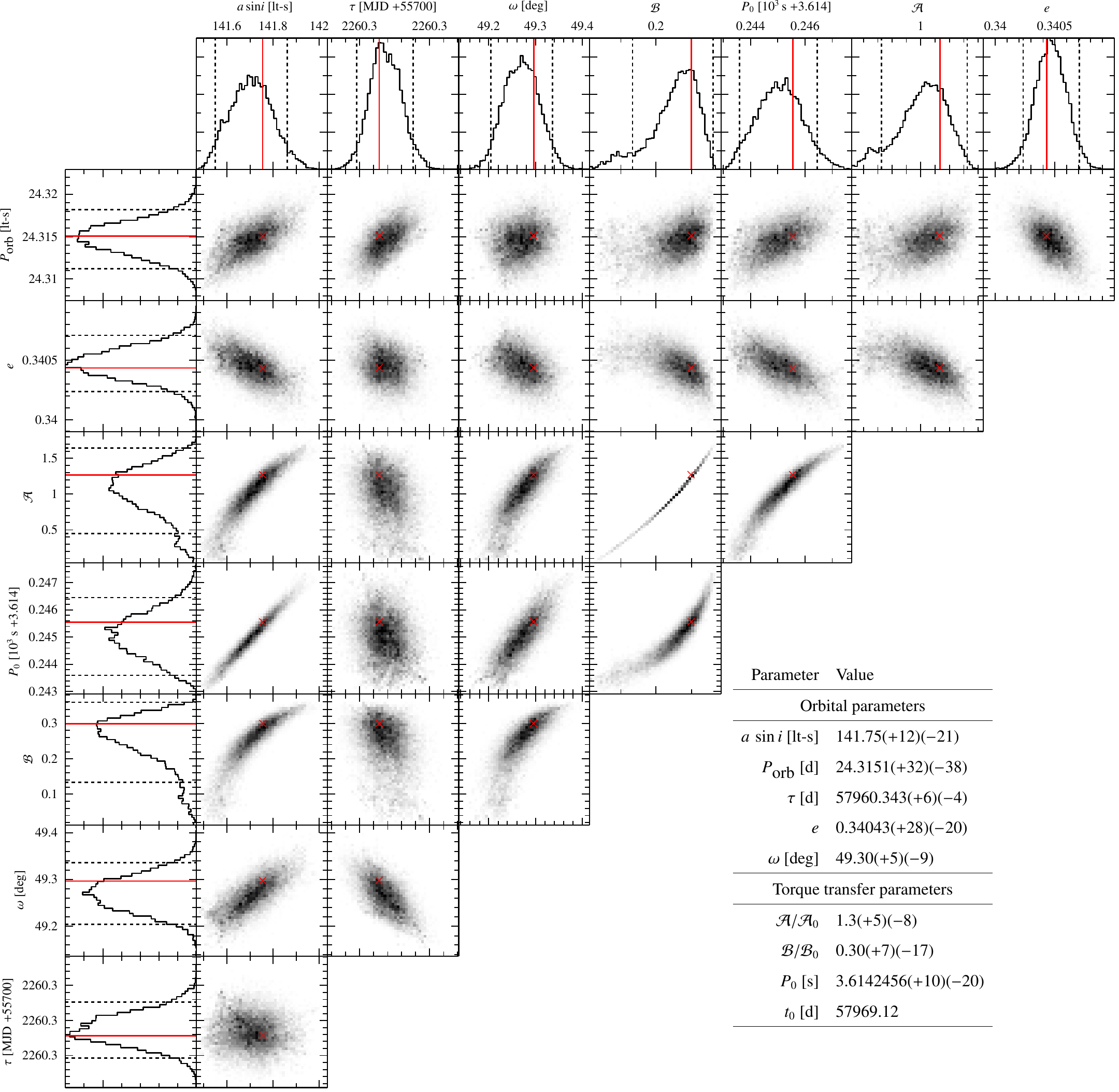}
\caption{Resulting parameter distribution of W$_\tn{III}$ for the 2017
outburst of 4U\,0115. See Fig.~\ref{fig:mcmc-1}.}
\label{fig:mcmc-3}
\end{figure*}

\begin{figure*}
\includegraphics[width=\hsize]{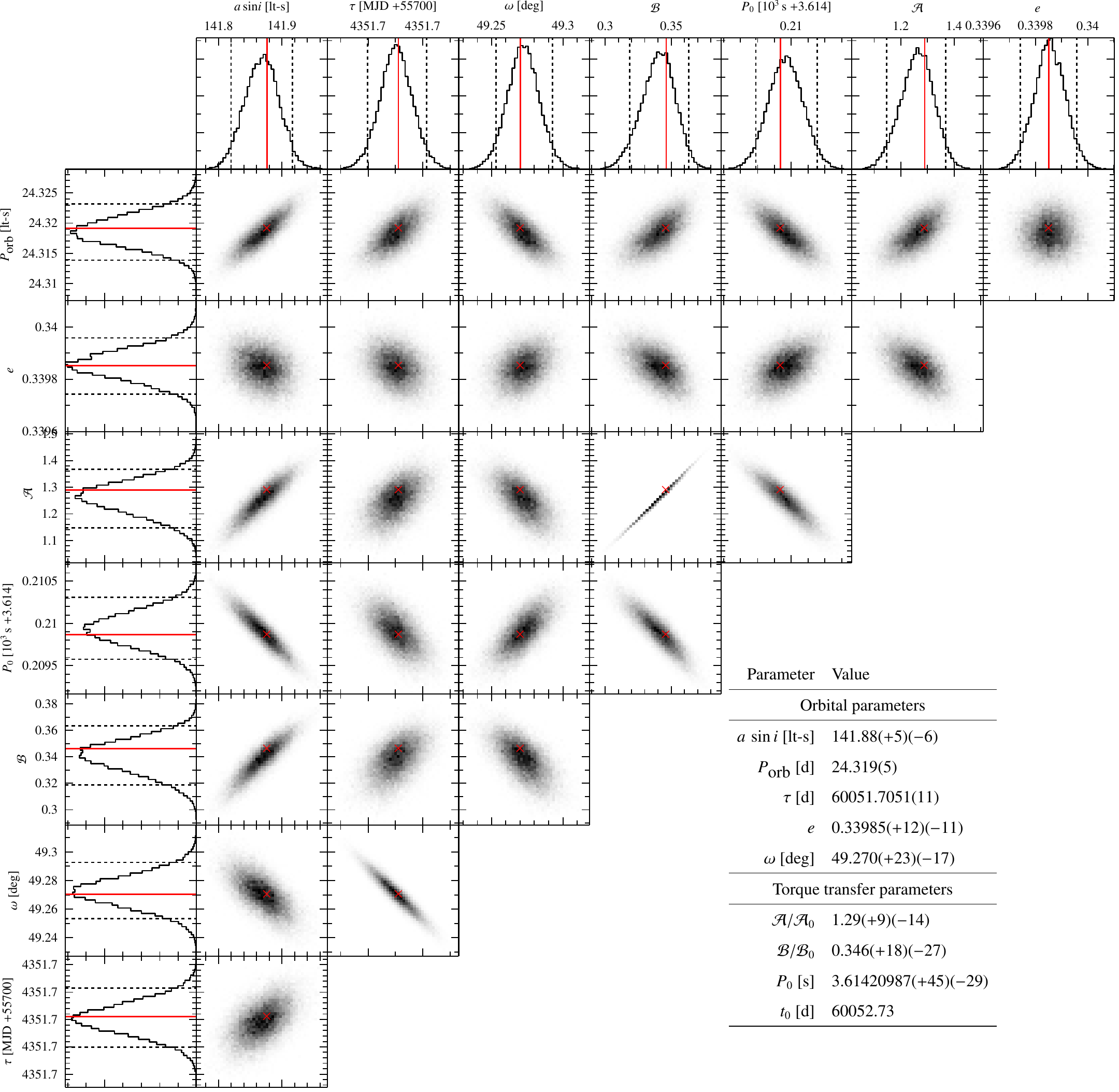}
\caption{Resulting parameter distribution of W$_\tn{III}$ for the 2023
outburst of 4U\,0115. See Fig.~\ref{fig:mcmc-1}.}
\label{fig:mcmc-4}
\end{figure*}

Figs.~\ref{fig:mcmc-1} to \ref{fig:mcmc-4} show only the results of the
model of \citet[III]{wang1995a} for all four outbursts. The distributions obtained for the other models are
almost identical except for the expected differences in $\mathcal{A}$ and $\mathcal{B}$.

\end{appendix}

\end{document}